\documentclass[12pt]{article}
\usepackage{amssymb}
\usepackage{amsmath,bm}
\usepackage{graphics,mathrsfs}
\usepackage{graphicx,epsfig}
\usepackage{subfigure}
\usepackage{placeins}
\usepackage{indentfirst}
\usepackage{setspace}
\usepackage{enumerate}
\usepackage{enumitem}
\usepackage{caption}
\usepackage{varioref}
\usepackage{color}
\usepackage[numbers]{natbib}
\usepackage{epstopdf}
\usepackage{booktabs,tabularx}
\usepackage{amsthm}
\makeatletter \oddsidemargin  -.1in \evensidemargin -.1in
\begin{document}
	%\doublespacing
	\newcommand{\bea}{\begin{eqnarray}}
		\newcommand{\eea}{\end{eqnarray}}
	\newcommand{\nn}{\nonumber}
	\newcommand{\bee}{\begin{eqnarray*}}
		\newcommand{\eee}{\end{eqnarray*}}
	\newcommand{\lb}{\label}
	\newcommand{\nii}{\noindent}
	\newcommand{\ii}{\indent}
	\newtheorem{thm}{Theorem}[section]
	\newtheorem{example}{Example}[section]
	\newtheorem{cor}{Corollary}[section]
	\newtheorem{definition}{Definition}[section]
	\newtheorem{lem}{Lemma}[section]
	\newtheorem{rem}{Remark}[section]
	\newtheorem{proposition}{Proposition}[section]
	\numberwithin{equation}{section}
	\renewcommand{\theequation}{\thesection.\arabic{equation}}
	\renewcommand\bibfont{\fontsize{10}{12}\selectfont}
	\setlength{\bibsep}{0.0pt}
	\title{\bf Statistical inference for Gumbel Type-II distribution under simple step-stress life test using Type-II censoring*}
%	\author{Subhankar Dutta$^{1}$, Farha Sultana$^{2}$, Suchandan Kayal$^{3}$\thanks{Corresponding author : Suchandan Kayal ( kayals@nitrkl.ac.in,~suchandan.kayal@gmail.com)}
	\author{ Subhankar Dutta$^1$\thanks {Email address(corresponding author): subhankar.dta@gmail.com}~, Farha Sultana$^2$\thanks{Email adress : farhasultana18@gmail.com}~, Suchandan Kayal$^1$\thanks {Email address 
		kayals@nitrkl.ac.in,~suchandan.kayal@gmail.com~~~ *It has been published in Iranian Journal of Science, https://doi.org/10.1007/s40995-022-01394-3}}
	%\doublespacing
	
	\maketitle
	\noindent {\it $^{1,3}$Department of Mathematics, National Institute of
	Technology, Rourkela-769008, India.} \\
	{\it $^{2}$ Department of Mathematics, Indian Institute of Information Technology, Guwahati-781015, India}
	\date{}
	\maketitle
	%	\noindent {\it $^{1}$Department of Mathematics, National Institute of
	%		Technology, Rourkela-769008, India.} \\
	%	{\it $^{2}$ Department of Mathematics, Indian Institute of Information Technology, Guwahati-781015, India.}
	%{\it $^{3}$ Applied Statistics Unit, Indian Statistical Institute, Kolkata-700108, India.}
	%\maketitle
	%\date{}
	%\maketitle 
	\begin{center}
		\textbf{Abstract}
	\end{center}
	\noindent In this paper, we focus on the parametric inference based on the Tampered Random Variable (TRV) model for simple step-stress life testing (SSLT) using Type-II censored data. The baseline lifetime of the experimental units, under normal stress conditions, follows  the Gumbel Type-II distribution with $\alpha$ and $\lambda$ being the shape and scale parameters, respectively. Maximum likelihood estimator (MLE) and Bayes estimator of the model parameters are derived based on Type-II censored samples. We obtain asymptotic intervals of the unknown parameters using the observed Fisher information matrix. Bayes estimators are obtained using Markov Chain Monte Carlo (MCMC) method under  squared error loss and LINEX loss functions. We also construct highest posterior density (HPD) intervals of the unknown model parameters.  Extensive simulation studies are performed to investigate the finite sample properties of the proposed estimators. Three different optimality criteria have been considered to determine the optimal censoring plans. Finally, the methods are illustrated with the analysis of two real data sets.\\
	\\
	\textbf{Keywords:} Step-stress life testing; Tampered random variable model,  Bayesian Analysis, MCMC method, Metropolis-Hastings algorithm, Highest posterior density credible interval, Optimality.
	
	\section{Introduction}
	
	In reliability analysis, there are many problems with life testing experiments which require a
	long time to acquire the test data such as for highly reliable products at the specified use condition. To get enough information about these products' lifespan characteristics, the typical life test of these products under normal working settings is too time-consuming and expensive. This is a significant concern for the high-tech industries since it could cause significant delays in the release of newly developed or improved products, resulting in lost commercial opportunities and market share losses. Accelerated life tests (ALT) are used to address this issue since they allow for the collection of more failure data in a shorter amount of time by subjecting test units to higher stress levels (temperature, pressure, voltage, vibration, etc.) than usual. The constant-stress accelerated life test (CSALT) enables the experimenters to divide the items into various groups, with each group being tested at various stress levels. However, in CSALT, the experiment may last excessively long because there is frequently a large scatter in failure times under low stress levels. To overcome such issues a special class of ALTs, known as step-stress life testing (SSLT) is introduced. As a result, the step-stress accelerated life test (SSALT) is suggested. With the SSLT, the stress levels can be altered during the experiment at predetermined time points or after a predetermined number of failures. In such life testing experiment with two stress levels $s_1$, $s_2$ (say), and $n$ identical units are placed on test initially under normal stress level $s_1$.  The stress level is changed from  $s_1$ to $s_2$, at the pre-fixed time $\tau$, known as the tampering time. There may  be more stress levels and corresponding to each stress change there would be more than one tampering time points, we call it multiple step-stress life testing. If there are only two stress levels with one  tampering time, then it is known as the simple step-stress life testing, or simple SSLT. The lifetime distribution under the initial normal stress level is termed as the baseline lifetime distribution.
	Some key references on the ALT model are  Nelson \cite{nelson1980}, Bhattacharyya and Soejoeti \cite{bhattacharyya1989}, Madi \cite{madi1993}, Bai and Chung \cite{bai1992}.
	\par Modern products and technologies are getting more sophisticated and reliable due to ongoing advancements in engineering technology and production techniques. Industries like electrical gadgets, computer equipment, vehicle parts, and others have quite high mean times to failure. It is generally impractical, expensive, and time-consuming to conduct life tests under typical operating conditions. In order to end the life testing experiment in a controlled manner before all the items fail, censoring is a standard statistical strategy. There are many situations in life testing and reliability experiments in which the experiment stops earlier (before all units fail) and all remaining units at this time point are censored at once. The most commonly used censoring schemes are Type-I and Type-II censoring schemes. 
	%Briefly, these can be described as follows. Consider $n$ units are  placed under observations  on a particularexperiment. In the conventional Type-I censoring scheme, the experiment continues upto a pre-specified time $\eta$, which is pre-fixed. The main drawback of this censoring scheme is one may get very few failures or in worst case no failure till time $\eta$. Therefore,  estimating the unknown parameters can not be done efficiently. On the other hand,
	In the conventional Type-II censoring scheme the experiment continues until a pre-specified number $r$, (say), of failures ($r\leq n$) occurs. Therefore, the Type-II censoring always ensures $r$ number of failures during the life testing experiment. 
	% Here, also the major drawback is the experimental time. If one uses a highly reliable product for the life testing experiment then it will take more time to get a pre-specified number of failures and which leads to a cost constraint for the experimenter.
	
	For a simple SSLT, the cumulative exposure (CE) model has been widely used in statistical literature. Sedyakin \cite{sedyakin1966one} proposed this CE model and then it has been extensively studied by Nelson \cite{nelson1980}. Under two different stress conditions, if $F_1(\cdot)$ and $F_2(\cdot)$ represent two different distributions, then the lifetime distribution under CE model can be expressed as 
	\begin{align}
		F_{CE}(T)=\begin{cases}
			F_{1}(T), ~~~~~~~~~~~~~\mbox{if}~~T\leq \tau,\\
			F_{2}(T-\tau+h),~~\mbox{if}~~ T > \tau,
		\end{cases}
	\end{align}
	where $h$ can be determined by solving the equation $F_{2}(h)=F_{1}(\tau)$. 
	
	\par Bhattacharyya and Soejoeti \cite{bhattacharyya1989tampered} proposed another SSLT model, named as tampered failure rate (TFR) model which has gained a lot of attention in recent years. If $\lambda_{TFR}(t)$ denotes the failure time of the overall lifetime distribution under SSLT, then the TFR model can be expressed as 
	\begin{align}
		\lambda_{TFR}(t)= \begin{cases}
			\lambda_1(t), ~~~~~\mbox{if}~~t \leq \tau,\\
			\alpha \lambda_1(t), ~~~\mbox{if}~~ t>\tau,
		\end{cases}
	\end{align}
	where $\lambda_1(t)$ is the initial failure rate in normal stress condition and $\alpha$ is an unknown factor (usually greater than $1$).
	
	\par Goel \cite{goel1971} first introduced the tampered random variable (TRV)
	modeling in the context of a simple SSLT (see also DeGroot and Goel \cite{degroot1979}), which assumes
	that the effect of change of the stress level at time $\tau$ is equivalent to changing the remaining life of the experimental unit by an unknown positive factor, say $\beta$ (usually,
	less than 1). Let $T$ be the random variable representing the baseline lifetime under normal stress condition. Then, the overall lifetime, denoted by the random variable $T_{TRV}$, is defined
	as
	\begin{eqnarray*}
		T_{TRV} = \left\{
		\begin{array}{ll}
			\displaystyle T,&~ \mbox{if}~0 < T \leq \tau, \\ \nonumber
			\displaystyle \tau+\beta  (T-\tau),& ~\mbox{if}~ T>\tau,
		\end{array} \right.
	\end{eqnarray*}
	where the scale factor $\beta$, called the tampering coefficient, depends on both the stress levels $s_1$, $s_2$ and
	possibly on $\tau$ as well. The time point  $\tau$ is called the tampering time. Note that all these three above discussed models are equivalent if the baseline lifetime follows exponential distribution. For more details one can see Sultana and Dewanji \cite{sultana2021tampered}. In literature many authors considered TRV model for estimating different lifetime distributions. Abdel-Ghaly et al. \cite{abdel2002} considered the estimation problem of the Weibull distribution in ALT. Maximum likelihood estimators (MLEs) are obtained for the distribution parameters and the acceleration factor in both Type-I and Type-II censored samples. The modified quasilinearization method is used to solve the nonlinear maximum likelihood equations. Also, the confidence intervals of the estimators are obtained. Wang et al. \cite{wang2012} studied the estimation of the parameters of the Weibull distribution in step-stress ALT under multiply censored data. The MLEs are used to obtain the parameters of the Weibull distribution and the acceleration factor under multiply censored data. Additionally, the confidence intervals for the estimators are also obtained. Ismail \cite{ismail2014} obtained the MLEs of Weibull distribution parameters and the acceleration factor under adaptive Type-II progressively hybrid censored data. The method has been extended for an adaptive Type-I progressive hybrid censored data by Ismail \cite{ismail2016}. El-Sagheer et al. \cite{el2019inferences} discussed point and interval estimates of the parameters for Weibull-exponential distribution using partially accelerated step-stress model under progressive Type-II censoring. Amleh and Raqab \cite{amleh2021inference} obtained statistical inference for Lomax distribution based on simple step-stress under Type-II censoring.  Nassar et al. \cite{nassar2021bayesian} discussed expected Bayes estimation using simple step-stress under Type-II censoring scheme. Ramzan et al. \cite{ramzan2022bayesian} discussed classical and Bayesian estimation using simple SSLT based on TRV model for modified Weibull distribution under Type-I censoring scheme.  
	\par Let us consider the baseline lifetime $T$ follows the Gumbel Type-II  distribution and the corresponding probability density function (PDF)
	and cumulative distribution function (CDF) are, respectively,
	\begin{eqnarray}
		f_T(t)= \alpha \lambda t^{-(\alpha +1)}e^{-\lambda t^{-\alpha}}, \quad t>0,~~\alpha>0, ~~\lambda>0,
	\end{eqnarray}
	and
	\begin{eqnarray}
		F_T(t)= e^{-\lambda t^{-\alpha}}, \quad t>0,
	\end{eqnarray}
	where $\alpha$, $\lambda $ are the shape and scale parameters, respectively. The hazard rate function of the Gumbel Type-II distribution is decreasing or upside-down bathtub (UTB) shape depends on the parameters values. Due to these shapes of the hazard rate function, the Gumbel Type-II distribution is very flexible to model meteorological phenomena such as floods, earthquakes, and natural disasters, also in medical and epidemiological applications. In recent years many authors have studied statistical properties of the estimators of the model parameters of the Gumbel Type-II distribution. Abbas et al. \cite{abbas2013bayesian} discussed the Bayesian estimation of the model parameters of the Gumbel Type-II distribution. Then E-Bayesian estimation of the unknown model shape parameter has been studied by Reyad and Ahmed \cite{reyad2015bayesian}. Sindhu et al. \cite{sindhu2016study} obtained the Bayes estimates and corresponding risk of the model parameters based on left censored data.
	\par When failure data are acquired through a life test, statistical inference of the product based failure data is a crucial issue. In view of the above discussed concerns, statistical techniques, and time restrictions in many tests, we consider the statistical inference of the Gumbel Type-II distribution under simple SSLT based on Type-II censoring. To the best of our knowledge, this problem has not been studied yet. The maximum likelihood and Bayesian estimation techniques are considered in the inferential aspects. Additionally, we present a set of recommendations for selecting the most effective estimating technique to estimate the unknown model parameters under the SSLT model, which we believe would be of great interest to applied statisticians and reliability engineers. The objective of the optimization of this model is to identify the censoring plan which leads to the most precise estimation of criteria. Under this consideration, three different optimality criteria have been considered based on the observed Fisher information matrix. The rest of the article is organized as follows. In Section 2, we introduce the TRV modelling under simple SSLT and derive the corresponding CDF and PDF for Gumbel Type-II baseline lifetimes. Also, the MLEs of the unknown parameters,  $\alpha$, $\lambda$, and $\beta$ are derived using Type-II censored samples.
	We also construct asymptotic confidence intervals based on the observed Fisher information matrix and bootstrap confidence intervals of unknown parameters. Further, Bayes estimates are obtained under the squared error loss as well as LINEX loss functions in Section 3. We compute these estimates using the MH-algorithm of MCMC method. The HPD credible intervals of unknown parameters are discussed as well. Section 4 presents some simulation studies to investigate the finite sample properties of the MLEs. In Section 5, optimal censoring schemes based on different optimality criteria have been investigated.  We illustrate the proposed methods through the analysis of two real life data sets in  Section 6 while Section 7 ends with some concluding remarks.
	\section{Model Description and MLEs}
	%\noindent In this section we will discuss about the TRV modeling under Type-II censoring. Different types of scenarios of the failure data are described below.
	%\subsection{Model Description}
	\noindent In a simple SSLT model, let us consider that $n$ number of experimental units are placed with initial stress $s_{1}$. After a prefixed time $\tau$, the initial stress level is changed from $s_1$ to $s_{2}$. The experiment will be terminated when $r^{th}$ failure occurs, where $r$ is a pre-fixed integer. Therefore, the time of failures $t_{1:n} < t_{2:n} < ... <t_{r:n} $, are denoted as the observed data. The following are all possible  types of data we can get from the Type-II censoring under simple SSLT: \\
	\textbf{Case-I :}~~ $t_{1:n} < t_{2:n} < \cdots < t_{r:n} < \tau$, \\
	\textbf{Case-II :}~ $t_{1:n} < t_{2:n} < \cdots < t_{N:n} < \tau < t_{N+1:n} < t_{N+2:n} < \cdots < t_{r:n}$,\\
	\textbf{Case-III :} $\tau < t_{1:n} < t_{2:n} < \cdots < t_{r:n}$,\\
	where N is the number of failures at normal stress level $s_{1}$. In particular, for Case-I, $N=r$ and for Case-III, $N=0$. Basically Case-I and Case-III are the special cases of Case-II, thereafter we will only focus on Case-II in the remaining part of this paper.
	\par Let us assume that the baseline lifetime follows the Gumbel Type-II distribution with $\alpha$, $\lambda$ are shape and scale parameters, respectively. Also, assume that the experimental units are independent and identically distributed (i.i.d.) in the life testing experiment. Now, under the assumption of TRV model, the CDF $F_{TRV}(.)$, of $T_{TRV}$ is given by
	\begin{align}
		F_{{TRV}}(t) & = \begin{cases}
			F_T(t), & \text{if }0\leq t<\tau,\\
			F_T\left(\tau+\frac{t-\tau}{\beta}\right), & \text{if }t\geq\tau,
		\end{cases}~= \begin{cases}
			e^{-\lambda t^{-\alpha}}, & \text{if } 0 \leq t < \tau, \\
			e^{-\lambda (\tau + \frac{t-\tau}{\beta})^{-\alpha}}, & \text{if } t \geq \tau.
		\end{cases} \label{1.1}
	\end{align}
	The corresponding PDF is given by
	\begin{align}
		f_{{TRV}}(t)& = \begin{cases}
			f_T(t), & \text{if }0\leq t<\tau,\\
			\frac{1}{\beta}f_T\left(\tau+\frac{t-\tau}{\beta}\right), & \text{if }t\geq\tau,
		\end{cases}= \begin{cases}
			\alpha \lambda t^{-(\alpha +1)}e^{-\lambda t^{-\alpha}}, & \text{if } 0 \leq t < \tau, \\
			\frac{\alpha \lambda}{\beta} (\tau + \frac{t-\tau}{\beta})^{-(\alpha +1)}e^{-\lambda (\tau + \frac{t-\tau}{\beta})^{-\alpha}}, & \text{if } t \geq \tau.
		\end{cases} \label{1.2}
	\end{align}
	Next we will discuss about the MLEs of the unknown model parameters.
	\subsection{Maximum Likelihood Estimation}
	In this section, we will determine MLEs for the unknown model parameters under Type-II censored sample using TRV modeling based on simple SSLT. Let $T_{1:n},T_{2:n}, \cdots,T_{n:n}$ be the random sample of size $n$ from the Gumbel Type-II distribution described in (\ref{1.1}) with the unknown model parameters $\alpha$, $\lambda$, and $\beta$.
	Therefore, the likelihood function for TRV modeling under Type-II censoring can be written as
	\begin{eqnarray*}
		L(\alpha,\lambda,\beta|data)\propto \prod_{i=1}^{N}f_{T}(t_{i:n}) \prod_{i=N+1}^{r}\bigg(\frac{1}{\beta}f_{T}\left(\tau+\frac{t_{i:n}-\tau}{\beta}\right)\bigg)\left[1-F_{T}\left(\tau+\frac{t_{r:n}-\tau}{\beta}\right)\right]^{n-r}.
	\end{eqnarray*}
	Thus, the likelihood function for the Gumbel Type-II distribution can be written as
	\begin{align}
		\nonumber L(\alpha,\lambda,\beta|data) \propto~~ & {\alpha}^r {\lambda}^r {\beta}^{r-N} \prod_{i=1}^{N}{t_{i:n}}^{-(\alpha +1)} e^{-\lambda {t_{i}}^{-\alpha}} \prod_{i=N+1}^{r} \bigg(\tau + \frac{t_{i:n}-\tau}{\beta}\bigg)^{-(\alpha +1)}\\
		& \times e^{-\lambda \sum_{i=N+1}^{r}\bigg(\tau + \frac{t_{\textcolor{blue}{i:n}}-\tau}{\beta}\bigg)^{-\alpha}} \bigg[1-e^{-\lambda \left(\tau + \frac{t_{r:n}-\tau}{\beta}\right)^{-\alpha}}\bigg]^{n-r}.
	\end{align}
	Therefore, the log-likelihood function can be written as
	\begin{align}
		\nonumber l(\alpha,\lambda,\beta)=~~& r\log \alpha + r\log \lambda +(r-N) \log \beta-(\alpha+1)\bigg( \sum_{i=1}^{N} \log t_{i} + \sum_{i=N+1}^{r}\log z_{i}\bigg) \\
		&  -\lambda \bigg(\sum_{i=1}^{N} {t_{i}^{-\alpha}}+\sum_{i=N+1}^{r} {z_{i}^{-\alpha}}\bigg)  + (n-r) \log \bigg(1-e^{-\lambda {z_{r}^{-\alpha}}}\bigg), \label{1.4}
	\end{align}
	where, $z_{i}=~\tau + \frac{t_{i}-\tau}{\beta} $ and $z_{r}=~ \tau + \frac{t_{r}-\tau}{\beta}$. Note that, for notational simplicity, in the rest of the paper we write $t_{i:n}$ and  $t_{r:n}$ as $t_i$ and $t_{r},$ respectively.
	
	\par Now, taking partial derivatives of $l(\alpha,\lambda,\beta)$ with respect to unknown parameters we get likelihood equations as given below
	
	\begin{align}
		\nonumber \frac{\partial l}{\partial \alpha} =~~& \frac{r}{\alpha} -\bigg(\sum_{i=1}^{N}\log t_{i}+\sum_{i=N+1}^{r} \log z_{i}\bigg) + \lambda \bigg(\sum_{i=1}^{N} {t_{i}^{-\alpha}} \log t_{i}  + \sum_{i=N+1}^{r} {z_{i}^{-\alpha}} \log z_{i}\bigg) \\
		& -(n-r)\frac{\lambda {z_{r}^{-\alpha}}\log z_{r}e^{-\lambda {z_{r}^{-\alpha}}}}{(1-e^{-\lambda {z_{r}^{-\alpha}}})} =0, \label{1.5}\\
		\frac{\partial l}{\partial \lambda} =~~& \frac{r}{\lambda} -\sum_{i=1}^{N} {t_{i}^{-\alpha}} -\sum_{i=N+1}^{r} {z_{i}^{-\alpha}} + (n-r)\frac{{z_{r}^{-\alpha}}}{(e^{\lambda {z_{r}^{-\alpha}}}-1)} =0, \label{1.6}\\
		\nonumber \mbox{and}~~~~~~~~~~~~~~&\\
		\nonumber \frac{\partial l}{\partial \beta} =~~& \frac{r-N}{\beta} -(\alpha +1) \sum_{i=N+1}^{r}\frac{{z_{i}^{\prime}}}{z_{i}} -\alpha\lambda \sum_{i=N+1}^{r} {z_{i}^{-(\alpha+1)}} {z_{i}^{\prime}} \\
		&-(n-r) \frac{\alpha\lambda {z_{r}^{-(\alpha+1)}}{z_{r}^{\prime}} }{(e^{\lambda {z_{r}^{-\alpha}}}-1)}=0, \label{1.7}
	\end{align}
	where ${z_{i}^{\prime}} = \frac{\tau -t_{i}}{{\beta}^2}$ and  ${z_{r}^{\prime}} = \frac{\tau -t_{r}}{{\beta}^2}$.\\
	As the likelihood equations are in implicit form of the unknown parameters, thus we cannot solve it explicitly to determine the MLEs of $\alpha$, $\lambda$, and $\beta$ as $\widehat{\alpha}$, $\widehat{\lambda}$, and $\widehat{\beta}$. Therefore we solve (\ref{1.5}), (\ref{1.6}), and (\ref{1.7}) numerically by using some numerical method, such as Newton-Raphson.
	
	\subsection{Approximate Confidence Intervals}
	In this section we want to construct the asymptotic confidence intervals for the unknown model parameters. To obtain $100(1-\gamma) \%$ confidence intervals with $\gamma$ significance level of the unknown parameters of the Gumbel Type-II distribution under simple SSLT, we have to calculate the asymptotic variance-covariance matrix. Using asymptotic normality properties of MLEs of the parameters, an asymptotic variance-covariance matrix can be obtained. In doing so, the variance of $\widehat{\alpha}$, $\widehat{\lambda}$, and $\widehat{\beta}$ are required. These can be obtained from the diagonal elements of the inverse of the observed Fisher information matrix, $\widehat{I}^{-1}(\widehat{\alpha},\widehat{\lambda},\widehat{\beta})$, where \\
	\begin{align}\label{1.8}
		\widehat{I}(\widehat{\alpha},\widehat{\lambda},\widehat{\beta})= {\begin{bmatrix}
				-l_{11} & -l_{12} & -l_{13}\\
				-l_{12} & -l_{22} & -l_{23}\\
				-l_{13} & -l_{23} & -l_{33}\\
		\end{bmatrix}}_{({\alpha},~{\lambda},~{\beta})=(\widehat{\alpha},~\widehat{\lambda},~\widehat{\beta})},
	\end{align}
	and $l_{ij}= \frac{\partial^2 logL}{\partial \theta_{i}\partial \theta_{j}}$ for $i,j=1,2,3,$ where  $\Theta=(\theta_{1},\theta_{2},\theta_{3})= (\alpha,\lambda,\beta)$. So here,\\
	\begin{eqnarray}
		\nonumber l_{11}&=& -\frac{r}{{\alpha}^2} -\lambda \sum_{i=1}^{N}\frac{(\log {t_{i}})^2}{{t_{i}^{\alpha}}}- \lambda \sum_{i=N+1}^{r}\frac{(\log {z_{i}})^2}{{z_{i}^{\alpha}}} -(n-r)\lambda {z_{r}^{-\alpha}} (\log {z_{r}})^2 \\
		& \nonumber \times & \frac{ (1-e^{\lambda {z_{r}^{-\alpha}}}+\lambda {z_{r}^{-\alpha}}e^{\lambda {z_{r}^{-\alpha}}})}{(e^{\lambda {z_{r}^{-\alpha}}}-1)^2}, \\
		\nonumber l_{12}=l_{21}&=& \sum_{i=1}^{N}\frac{\log {t_{i}}}{{t_{i}}^{\alpha}} + \sum_{i=1}^{N}\frac{\log {z_{i}}}{{z_{i}}^{\alpha}} -(n-r){z_r^{-\alpha}} \log z_r \frac{(e^{\lambda {z_{r}^{-\alpha}}}-1-\lambda z_r^{-\alpha} e^{\lambda {z_{r}^{-\alpha}}})}{(e^{\lambda {z_{r}^{-\alpha}}}-1)^2}, \\
		\nonumber l_{13}=l_{31}&=&  - (n-r) \frac{\lambda {z_{r}^{\prime}}\bigg({z_{r}^{\alpha}}(e^{\lambda {z_{r}^{-\alpha}} }-1)+\alpha \bigg({z_{r}^{\alpha}}+e^{\lambda {z_{r}^{-\alpha}}}(\lambda-{z_{r}^{\alpha}})\bigg)\log z_{r} \bigg) }{{z_{r}^{2\alpha +1}}(e^{\lambda {z_{r}^{-\alpha}}}-1)^2}\\
		& & \nonumber -\sum_{i=N+1}^{r} \frac{{z_{i}^{\prime}}}{z_{i}}
		-\lambda\sum_{i=N+1}^{r}  {z_{i}^{-\alpha}}\frac{{z_{i}^{\prime}}}{z_{i}} (\alpha\log z_{i} -1),\\
		\nonumber l_{22}&=& -\frac{r}{{\lambda}^2}-(n-r)\frac{{z_{r}^{-2\alpha}}e^{\lambda {z_{r}^{-\alpha}} }}{(e^{\lambda {z_{r}^{-\alpha}} }-1)^2},\\
		\nonumber l_{23}=l_{32}&=& \sum_{i=N+1}^{r} \alpha {z_{i}^{-\alpha}} \frac{{z_{i}^{\prime}}}{z_{i}}- (n-r) \frac{{z_{r}^{\prime}}}{z_{r}} \frac{{z_{r}^{-2\alpha}}}{(e^{\lambda {z_{r}^{-\alpha}}}-1)}\bigg(e^{\lambda{z_{r}^{-\alpha}}}({z_{r}^{\alpha}}-\lambda)-{z_{r}^{\alpha}}\bigg),
	\end{eqnarray}
	and
	\begin{eqnarray}
		\nonumber l_{33}&=& -\frac{r-N}{{\beta}^2} + (\alpha+1)\sum_{i=1}^{r-N}\bigg(\frac{{z_{i}^{\prime}}}{z_{i}} \bigg[\frac{{z_{i}^{\prime}}}{z_{i}}+\frac{2}{\beta}\bigg] \bigg) -\alpha \lambda \sum_{i=N+1}^{r} \bigg({z_{i}^{-\alpha}}\frac{{z_{i}^{\prime}}}{z_{i}}\bigg[(\alpha+1)\frac{{z_{i}^{\prime}}}{z_{i}}-\frac{2}{\beta}\bigg]\bigg)\\
		& &\nonumber - \frac{\alpha \lambda}{{\beta}^2}\bigg(\frac{{z_{r}^{\prime}}}{z_{r}}\bigg)^2 \frac{{z_{r}^{-2\alpha}}}{(e^{\lambda{z_{r}^{-\alpha}}}-1)^2}\bigg[(\alpha-1)t_{r}{z_{r}^{\alpha}}-(2\beta+\alpha-1)\tau {z_{r}^{\alpha}}+\tau e^{\lambda{z_{r}^{-\alpha}}}\bigg((2\beta-1){z_{r}^{\alpha}}\\
		& &\nonumber +\alpha({z_{r}^{\alpha}}-\lambda)\bigg)+t_{r}e^{\lambda{z_{r}^{-\alpha}}}\bigg({z_{r}^{\alpha}}+\alpha(\lambda-{z_{r}^{\alpha}})\bigg)\bigg].
	\end{eqnarray}
	Then the $100(1-\gamma) \%$ approximate confidence intervals for $\alpha$, $\lambda$, and $\beta$ are given by
	\begin{align}
		\nonumber \bigg(\widehat{\alpha}~ \underline{+}~z_{\frac{\gamma}{2}}\sqrt{Var(\widehat{\alpha})}\bigg),~~ \bigg(\widehat{\lambda}~ \underline{+}~z_{\frac{\gamma}{2}}\sqrt{Var(\widehat{\lambda})}\bigg), ~~~\mbox{and}~~~ \bigg(\widehat{\beta}~ \underline{+}~z_{\frac{\gamma}{2}}\sqrt{Var(\widehat{\beta})}\bigg),
	\end{align}
	where $z_{\frac{\gamma}{2}}$ is the upper $\frac{\gamma}{2}$-th percentile of a standard normal distribution.\\
	
	\subsection{Bootstrap Confidence Intervals}
	To obtain confidence intervals, normal approximation works well when the sample size is large enough. In case of small sample size, bootstrap re-sampling technique provides more accurate result to obtain the confidence intervals. In this section, two commonly used parametric bootstrap such as bootstrap-$p$ (Boot-$p$) and bootstrap-$t$ (Boot-$t$) confidence intervals are derived for $\alpha$, $\lambda,$ and $\beta$. We recall that Efron \cite{efron1982jackknife} introduced Boot-$p$ interval as an alternative to approximate confidence interval and Hall \cite{hall1988theoretical} introduced another bootstrap method, called Boot-$t$ method. These two methods are non-parametric bootstrap methods. Later, Kundu et. al \cite{kundu2003analysis} proposed two parametric confidence intervals. To construct these two parametric bootstrap confidence intervals, the following steps can be used.
	
	\subsubsection*{Boot-$p$ confidence intervals}
	\noindent \textbf{Step 1:} Generate a simple step-stress sample $y=(y_1,\cdots,y_n)$ from the Gumbel Type-II distribution and compute the MLEs $\widehat{\alpha}$, $\widehat{\lambda}$, and $\widehat{\beta}$ under Type-II censoring scheme using TRV modeling.\\
	\textbf{Step 2:} Generate a bootstrap sample using the MLEs  $\widehat{\alpha}$, $\widehat{\lambda}$, and $\widehat{\beta}$ and calculate the bootstrap MLEs, denoted by  $\widehat{\alpha}^{*}$, $\widehat{\lambda}^{*}$, and $\widehat{\beta}^{*}$. \\
	\textbf{Step 3:} Repeat Step 2 up to $B$ times and obtain $\big(\widehat{\alpha}^{*(1)},\cdots, \widehat{\alpha}^{*(B)}\big)$, $\big(\widehat{\lambda}^{*(1)},\cdots, \widehat{\lambda}^{*(B)}\big)$, and $\big(\widehat{\beta}^{*(1)}, \cdots,\widehat{\beta}^{*(B)}\big)$.\\
	\textbf{Step 4:} Rearrange all these $\widehat{\alpha}^{*}$, $\widehat{\lambda}^{*},$ and $\widehat{\beta}^{*}$ in an ascending order, and we then obtain
	$\big(\widehat{\alpha}^{*[1]},\cdots, \widehat{\alpha}^{*[B]}\big)$, $\big(\widehat{\lambda}^{*[1])},\cdots, \widehat{\lambda}^{*[B]}\big)$,  and $\big(\widehat{\beta}^{*[1]}, \cdots,\widehat{\beta}^{*[B]}\big)$.\\	
	Now, the $100(1-\gamma)\%$ Boot-$p$ confidence intervals for $\alpha$, $\lambda,$ and $\beta$ are respectively given by
	\begin{align}
		\nonumber \bigg(\widehat{\alpha}^{*[\frac{\gamma B}{2}]}, \widehat{\alpha}^{*[B-\frac{\gamma B}{2}]}\bigg), ~~~ \bigg(\widehat{\lambda}^{*[\frac{\gamma B}{2}]}, \widehat{\lambda}^{*[B-\frac{\gamma B}{2}]}\bigg), ~~\mbox{and}~~\bigg(\widehat{\beta}^{*[\frac{\gamma B}{2}]}, \widehat{\beta}^{*[B-\frac{\gamma B}{2}]}\bigg).
	\end{align}

	\subsubsection*{Boot-$t$ confidence intervals}
	\noindent \textbf{Step 1:} Generate a simple step-stress sample $y=(y_1,\cdots,y_n)$ from the Gumbel Type-II distribution and compute the MLEs $\widehat{\alpha}$, $\widehat{\lambda}$, and $\widehat{\beta}$ under Type-II censoring scheme using TRV modeling.\\
	\textbf{Step 2:} Generate a bootstrap sample using the MLEs  $\widehat{\alpha}$, $\widehat{\lambda}$, and $\widehat{\beta}$ based on simple step-stress sample under Type-II censoring scheme and calculate the bootstrap MLEs  $\widehat{\alpha}^{*}$, $\widehat{\lambda}^{*}$, and $\widehat{\beta}^{*}$. \\
	\textbf{Step 3:} Compute $t$-statistics for $\alpha$, $\lambda$, and $\beta$ as $T_{\alpha}= \frac{\widehat{\alpha}^{*}-\widehat{\alpha}}{\sqrt{Var(\widehat{\alpha}^{*})}}$, $T_{\lambda}= \frac{\widehat{\lambda}^{*}-\widehat{\lambda}}{\sqrt{Var(\widehat{\lambda}^{*})}},$ and $T_{\beta}= \frac{\widehat{\beta}^{*}-\widehat{\beta}}{\sqrt{Var(\widehat{\beta}^{*})}}$, respectively.\\
	\textbf{Step 4:} Repeat Steps $2$ and $3$ up to $B$ number of times and we obtain $\big(T_{\alpha}^{(1)},\cdots,T_{\alpha}^{(B)}\big)$, $\big(T_{\lambda}^{(1)},\cdots,T_{\lambda}^{(B)}\big)$, and  $\big(T_{\beta}^{(1)},\cdots,T_{\beta}^{(B)}\big)$.\\
	\textbf{Step 5:} Rearrange these $T_{\alpha}$, $T_{\lambda}$, and $T_{\beta}$ in an ascending order and obtain $\big(T_{\alpha}^{[1]},\cdots,T_{\alpha}^{[B]}\big)$, $\big(T_{\lambda}^{[1]},\cdots,T_{\lambda}^{[B]}\big)$, and  $\big(T_{\beta}^{[1]},\cdots,T_{\beta}^{[B]}\big)$.\\
	Then, the two-sided $100(1-\gamma)\%$ Boot-$t$ confidence intervals for $\alpha$, $\lambda,$ and $\beta$ are respectively given by
	\begin{align}
		\nonumber \bigg(\widehat{\alpha}+T_{\alpha}^{[\frac{B\gamma}{2}]}, \widehat{\alpha}+T_{\alpha}^{[B-\frac{B\gamma}{2}]}\bigg), ~~\nonumber \bigg(\widehat{\lambda}+T_{\lambda}^{[\frac{B\gamma}{2}]}, \widehat{\lambda}+T_{\lambda}^{[B-\frac{B\gamma}{2}]}\bigg), ~~\mbox{and}~~\bigg(\widehat{\beta}+T_{\beta}^{[\frac{B\gamma}{2}]}, \widehat{\beta}+T_{\beta}^{[B-\frac{B\gamma}{2}]}\bigg).
	\end{align}
	
	\section{Bayesian Estimation}
	\noindent In this section, we will determine the Bayes estimates of the unknown parameters $\alpha$, $\lambda$, and $\beta$ based on different loss functions using Type-II censored data from the Gumble Type-II distribution. The most commonly used symmetric loss function is squared error loss (SEL) function and an asymmetric loss function is LINEX loss (LL) function. These loss functions are, respectively, defined as
	\begin{align}
		\nonumber L_{SE}\big(h(\theta),\widehat{h}(\theta)\big)=~& \big(h(\theta)-\widehat{h}(\theta)\big)^2, \\
		\nonumber \mbox{and}~~~~~~~~~~~~~~~~~~~~~~~~~~~~~~~~~~~~~~~~~~~~~~~~~~~~~~~~~~~~~~~~& \\
		\nonumber L_{LI}\big(h(\theta),\widehat{h}(\theta)\big)=~& e^{u\big(\widehat{h}(\theta)-h(\theta)\big)}-u\big(\widehat{h}(\theta)-h(\theta)\big)-1,~ u \neq 0,
	\end{align}
	where $\widehat{h}(\theta)$ is an estimate of a parametric function $h(\theta)$ and $u$ is a real number. All the parameters $\alpha$, $\lambda$, and $\beta$ are unknown. In this case, there doesn't exist any natural joint conjugate prior distribution. Thus, according to Kundu and Pradhan \cite{kundu2009bayesian} we assume independent priors for $\alpha$, $\lambda$, and $\beta$ as Gamma(a,b), Gamma(c,d), and Beta(p,q) distributions, respectively. We recall that $X\sim \mbox{Gamma}(a,b)$, if its PDF is given by
	\begin{eqnarray}
		f_{1}(x) \propto x^{a-1}e^{-bx},~x>0,~a,~b>0.
	\end{eqnarray}
	Further, if $X\sim \mbox{Beta}(p,q)$, then its PDF is
	\begin{eqnarray}
		f_{2}(x) \propto x^{p-1}(1-x)^{q-1},~0<x<1,~p,~q>0.
	\end{eqnarray}
	Now, the joint prior distribution of the unknown parameters is obtained as
	\begin{align}
		{\pi}^{*}(\alpha,\lambda,\beta) \propto {\alpha}^{a-1} e^{-b\alpha} {\lambda}^{c-1} e^{-d\lambda} {\beta}^{p-1} (1-\beta)^{q-1}, ~~ \alpha > 0,~ \lambda > 0,~ 0 < \beta < 1, \label{3.1}
	\end{align}
	where, $a, ~b,~ c, ~d,~ p,$ and $q$ are the hyper parameters. Note that the hyper parameters reflect the prior knowledge about the unknown parameters, and can take the value from any positive real numbers. After some calculations,  the joint posterior PDF of the unknown parameters $\alpha$, $\lambda$, and $\beta$ can be obtained as
	\begin{align}
		\nonumber \pi(\alpha,\lambda,\beta |data ) =&~  k \frac{{\alpha}^{r+a-1} {\lambda}^{r+c-1}}{{\beta}^{N+1-r-p}} (1-\beta)^{q-1} e^{-(b\alpha+d\lambda)} e^{-\big[(\alpha +1)\sum_{i=1}^{N}\log(t_{i}) +
			\lambda \sum_{i=1}^{N} {t_{i}}^{-\alpha}\big]} \\
		& \times e^{-\big[(\alpha +1) \sum_{i=N+1}^{r} \log z_i+ \lambda \sum_{i=N+1}^{r}z_i^{-\alpha} \big]}\big[1-e^{-\lambda {z_r^{-\alpha}}}\big]^{(n-r)}, \label{3.2}
	\end{align}
	where
	\begin{align}
		\nonumber k^{-1}= &\int_{0}^{1}\int_{0}^{\infty}\int_{0}^{\infty} \bigg[\frac{{\alpha}^{r+a-1} {\lambda}^{r+c-1}}{{\beta}^{N+1-r-p}} (1-\beta)^{q-1} e^{-(b\alpha+d\lambda)} e^{-\big[(\alpha +1)\sum_{i=1}^{N}\log(t_{i}) +
			\lambda \sum_{i=1}^{N} {t_{i}}^{-\alpha}\big]} \\
		\nonumber &\times e^{-\big[(\alpha +1) \sum_{i=N+1}^{r} \log z_i+ \lambda \sum_{i=N+1}^{r}z_i^{-\alpha} \big]}\big[1-e^{-\lambda {z_r^{-\alpha}}}\big]^{(n-r)}\bigg] d\alpha~d\lambda ~d\beta.
	\end{align}
	\noindent Under the loss functions SEL and LL, the Bayes estimates of $h(\alpha,\lambda,\beta)$ can be written, respectively, as $\widehat{h}_{SE}(\alpha,\lambda,\beta)$ and $\widehat{h}_{LI}(\alpha,\lambda,\beta)$ where \\
	\begin{align}
		\widehat{h}_{SE}(\alpha,\lambda,\beta) =& \int_{0}^{1}\int_{0}^{\infty}\int_{0}^{\infty} h(\alpha,\lambda,\beta) \pi(\alpha,\lambda,\beta | data )~ d\alpha~d\lambda~d\beta, \label{3.3}\\
		\nonumber\mbox{and}~~~~~~~~~~~~~~~~~~~~~~~~~~~~~~&\\
		\widehat{h}_{LI}(\alpha,\lambda,\beta) =& - \bigg(\frac{1}{u}\bigg) \log \bigg[\int_{0}^{1}\int_{0}^{\infty}\int_{0}^{\infty} e^{-uh(\alpha,\lambda,\beta)}\pi(\alpha,\lambda,\beta |data )~ d\alpha~d\lambda~d\beta\bigg]. \label{3.4}
	\end{align}
	Since $(\ref{3.3})$ and $(\ref{3.4})$ can not be solved explicitly, hence we use a numerical method to solve these equations.

	\subsection{MCMC Method}
	\noindent In this subsection, Markov Chain Monte Carlo (MCMC) method is adopted to enumerate the Bayes estimates of unknown parameters $\alpha$, $\lambda$, and $\beta$ under both the loss functions SEL and LL. In addition, HPD intervals are also composed by using the generated MCMC samples. From the posterior density function given by $(\ref{3.2})$, the conditional posterior densities can be written as
	\begin{align}
		\nonumber\pi_{1}(\alpha|\lambda,\beta,data) \propto &~ {\alpha}^{r+a-1} e^{-b\alpha} e^{-\big[(\alpha +1)\sum_{i=1}^{N}\log(t_{i}) +
			\lambda \sum_{i=1}^{N} {t_{i}}^{-\alpha}\big]} \big[1-e^{-\lambda {z_r^{-\alpha}}}\big]^{(n-r)}\\
		& \times e^{-\big[(\alpha +1) \sum_{i=N+1}^{r} \log z_i+ \lambda \sum_{i=N+1}^{r}z_i^{-\alpha}\big]},\label{3.5}\\
		\pi_{2}(\lambda|\alpha,\beta,data) \propto &~	{\lambda}^{r+c-1} e^{-d\lambda} e^{-
			\lambda \big[\sum_{i=1}^{N} {t_{i}}^{-\alpha}+ \sum_{i=N+1}^{r}z_i^{-\alpha}\big]} \big[1-e^{-\lambda {z_r^{-\alpha}}}\big]^{(n-r)},\label{3.6}\\
		\nonumber \mbox{and}~~~~~~~~~~~~~~~~~~~~~~~~~~~~~~&\\
		\pi_{3}(\beta|\alpha,\lambda,data) \propto &~ \frac{(1-\beta)^{q-1}}{{\beta}^{N+1-r-p}}~  e^{-\big[(\alpha +1) \sum_{i=N+1}^{r} \log z_i+ \lambda \sum_{i=N+1}^{r}z_i^{-\alpha}\big]} \big[1-e^{-\lambda {z_r^{-\alpha}}}\big]^{(n-r)}. \label{3.7}
	\end{align}
	
	The above density functions $\pi_{1}(\alpha|\lambda,\beta,data)$, $\pi_{2}(\lambda|\alpha,\beta,data)$, and $\pi_{3}(\beta|\alpha,\lambda,data)$ can not be written in the form of any well known distributions. Therefore, the MCMC samples can not be generated from these densities given in $(\ref{3.5})$, $(\ref{3.6})$, and $(\ref{3.7})$ directly. So, the Metropolis-Hastings algorithm is used to generate MCMC samples from the conditional densities. Then the Bayes estimates can be obtained by using the following steps:
	
	\noindent\textbf{Step 1:} Choose initial values as ${\alpha}^{(1)}= \widehat{\alpha}$, ${\lambda}^{(1)}=\widehat{\lambda}$, ${\beta}^{(1)}=\widehat{\beta}$,  and set $i=1$. \\
	\textbf{Step 2:} Generate ${\alpha}^{(i)}$, ${\lambda}^{(i)}$, and ${\beta}^{(i)}$ with normal distribution as ${\alpha}^{(i)}\sim N\big({\alpha}^{(i-1)},var(\widehat{\alpha})\big)$, ${\lambda}^{(i)}\sim N\big({\lambda}^{(i-1)},var(\widehat{\lambda})\big)$, and ${\beta}^{(i)}\sim N\big({\beta}^{(i-1)},var(\widehat{\beta})\big)$. \\
	\textbf{Step 3:} Compute ${\Omega}_{\alpha}= min\bigg(1,\frac{\pi_{1}\big({\alpha}^{(i)}|{\lambda}^{(i-1)}, {\beta}^{(i-1)}, data\big)}{\pi_{1}\big({\alpha}^{(i-1)}|{\lambda}^{(i-1)}, {\beta}^{(i-1)}, data\big)}\bigg)$, ${\Omega}_{\lambda}= min\bigg(1,\frac{\pi_{2}\big({\lambda}^{(i)}|{\alpha}^{(i-1)}, {\beta}^{(i-1)}, data\big)}{\pi_{2}\big({\lambda}^{(i-1)}|{\alpha}^{(i-1)}, {\beta}^{(i-1)}, data\big)}\bigg)$, and ${\Omega}_{\beta}= min\bigg(1,\frac{\pi_{3}\big({\beta}^{(i)}|{\alpha}^{(i-1)}, {\lambda}^{(i-1)}, data\big)}{\pi_{3}\big({\beta}^{(i-1)}|{\alpha}^{(i-1)}, {\lambda}^{(i-1)}, data\big)}\bigg)$. \\
	\textbf{Step 4:} Generate samples  for $\tau_{1}$, $\tau_{2}$, and $\tau_{3}$, where $\tau_{1} \sim Uniform(0,1)$, $\tau_{2} \sim Uniform(0,1)$, and $\tau_{3} \sim Uniform(0,1)$. \\
	\textbf{Step 5:} Set  \[ \left\{\begin{array}{lr}
		\alpha={\alpha}^{(i)}, & \mbox{if $\tau_{1} \leq {\Omega}_{\alpha}$; otherwise $\alpha={\alpha}^{(i-1)}$},\\
		\lambda={\lambda}^{(i)}, & \mbox{if $\tau_{2} \leq {\Omega}_{\lambda}$; otherwise $\lambda={\lambda}^{(i-1)}$},\\
		\beta={\beta}^{(i)}, & \mbox{if $\tau_{3} \leq {\Omega}_{\beta}$; otherwise $\beta={\beta}^{(i-1)}$}. \end{array}\right.\]\\
	\textbf{Step 6:} Set $i=i+1$. \\
	\textbf{Step 7:} Repeat steps $1$ to $6$, M times to get ${\alpha}^{(1)},\cdots,{\alpha}^{(M)}$; ${\lambda}^{(1)},\cdots,{\lambda}^{(M)}$, and ${\beta}^{(1)},\cdots,{\beta}^{(M)}$. \\
	Then, the Bayes estimates of $\alpha$, $\lambda$, and $\beta$ under SEL function are given as \\
	\begin{align}
		\nonumber	 \widehat{\alpha}_{SE}=\frac{1}{M}\sum_{i=1}^{M}{\alpha}^{(i)},~~~~\widehat{\lambda}_{SE}=\frac{1}{M}\sum_{i=1}^{M}{\lambda}^{(i)},~~~~\mbox{and}~~~~\widehat{\beta}_{SE}=\frac{1}{M}\sum_{i=1}^{M}{\beta}^{(i)}.
	\end{align}
	Further, the Bayes estimates of $\alpha$, $\lambda$, and $\beta$ under LL function are given as \\
	\begin{align}
		\nonumber	\widehat{\alpha}_{LI}=& -\frac{1}{u}\log \bigg(\frac{1}{M}\sum_{i=1}^{M}e^{-u{\alpha}^{(i)}}\bigg),~	\widehat{\lambda}_{LI}=& -\frac{1}{u}\log \bigg(\frac{1}{M}\sum_{i=1}^{M}e^{-u{\lambda}^{(i)}}\bigg),~
		\mbox{and}~ \widehat{\beta}_{LI}=& -\frac{1}{u}\log \bigg(\frac{1}{M}\sum_{i=1}^{M}e^{-u{\beta}^{(i)}}\bigg).
	\end{align}
	Furthermore, to construct $100(1-\gamma) \%$ HPD credible intervals for $\alpha$, $\lambda$, and $\beta$, we use the method given by Chen and Shao \cite{chen1999monte}. According to the method, the samples are re-arranged in increasing order and these are obtained as $({\alpha}^{[1]},\cdots,{\alpha}^{[M]})$, $({\lambda}^{[1]},\cdots,{\lambda}^{[M]})$, and $({\beta}^{[1]},\cdots,{\beta}^{[M]})$. Then, the $100(1-\gamma) \%$ HPD credible intervals are obtained as
	\begin{align}
		\nonumber \bigg({\alpha}^{[M\gamma/2]}, {\alpha}^{[M(1-\gamma/2)]}\bigg),~~~~\bigg({\lambda}^{[M\gamma/2]}, {\lambda}^{[M(1-\gamma/2)]}\bigg),~~~\mbox{and}~~~~\bigg({\beta}^{[M\gamma/2]}, {\beta}^{[M(1-\gamma/2)]}\bigg),
	\end{align}
	where $\gamma$ is the nominal significance level.
	
	\section{Simulation Study}
	In this section, a simulation study is carried out to compare the performance of different estimates of parameters for the Gumbel Type-II distribution under simple SSLT based on Type-II censoring. The performance of estimates are compared on the basis of the average estimate (AE) values and mean squared error (MSE). To observe the changes in the values of the parameters, the simulation study is constructed based on $10,000$ Type-II censored samples under simple SSLT. Three different choices of the sample size n= 50, 150, and 250 are taken to study the behavior of the estimates with change in the sample size. Also, corresponding to each sample size we choose some moderate values for $r$. We have also considered the Gumbel Type-II baseline lifetime distribution with two choices of the shape parameter $\alpha= 1, 1.5$ and the common scale parameter $0.75$. In the case of LL function, we consider the values of $u$ as $-0.05$ and $1$. Average bias (AB) and mean squared error (MSE) of the estimates are displayed in Tables $\ref{T1}$ and $\ref{T3}$. It is considered that, when the values of AB and MSE of an estimate become smaller then it performs better. Average width (AW) and coverage probabilities (CP) of asymptotic confidence interval (ACI), bootstrap confidence interval (BCI) and HPD credible intervals the corresponding parameters are tabulated in Tables $\ref{T2}$ and $\ref{T4}$. It is considered that, when AW of an interval becomes small then it performs better. In terms of CP, when CP becomes larger then this intervals performs better than other. To obtain Bayes estimates, different sets of hyperparameters have been chosen for $(a,b)$ as $(3,3)$ and $(3,2)$, for $(c,d)$ as $(3,4)$, and for $(p,q)$ as $(1,3)$ and $(7,10)$.  
	From Table $\ref{T1}$ and Table $\ref{T3}$, the following conclusions have been made:
	\begin{itemize}
		\item In most of the cases the values of ABs and MSEs decrease when values of $n$ increase. 
		\item For fixed value of $n$, ABs and MSEs of the estimators decrease when $r$ increases.
		\item For fixed $n$ and $r$, ABs and MSEs of the different estimates of the parameters decrease when $\tau$ increases. 
		\item For fixed $\tau$, $n$ and $r$, ABs and MSEs of the estimates decrease in most of the cases when $\beta$ increases. 
		\item Bayes estimates perform better than MLEs in terms of AB and MSE. Further, the Bayes estimate based on LL function (when $p=-0.05$) performs better than other estimates.
		\item When the value of $p$ tends to zero, Bayes estimate based on LL function performs as similar as SEL function based on AB and MSE. 
	\end{itemize}
	Now, from  Tables $\ref{T2}$ and $\ref{T4}$, the following conclusions can been made:
	\begin{itemize}
		\item For fixed $n$, the values of AW of the intervals decrease when $r$ increase\textcolor{blue}{s}.
		\item  For fixed $n$ and $r$, in most of the cases AW of the intervals decrease when $\beta$ increases.
		\item For fixed $n$ and $r$, in most of the cases AW of the intervals decrease when $\tau$ increases.
		\item In terms of AW, HPD credible interval performs better than another confidence intervals for any fixed $\tau$, $\beta$, $n$ and $r$. 
		\item In terms of CP, boot-p confidence interval performs better than another intervals for any fixed $\tau$, $\beta$, $n$ and $r$. 
	\end{itemize}
	From the above results, it can be summarized that Bayes estimates under LL function performs better than other estimates in terms of AB and MSE, and HPD credible intervals perform better than other confidence intervals in terms of AW.

	\begin{table}[htbp!]
		\begin{center}
			\caption{ABs and MSEs (in parantheses) of MLEs and Bayes estimates for  the  Gumbel Type-II  baseline  lifetime  distribution  with  scale parameter $\lambda=0.75$ and shape parameter $\alpha=1$ for different choices of $\beta$ and $\tau$ under simple SSLT. }
			\label{T1}
			\tabcolsep 7pt
			\small
			\scalebox{0.63}{
				\begin{tabular}{*{16}c*{15}{r@{}l}}
					\toprule
					\multicolumn{4}{c}{} &
					\multicolumn{3}{c}{MLE} & \multicolumn{3}{c}{SEL} & \multicolumn{3}{c}{LL~($u=-0.05$)}&\multicolumn{3}{c}{LL~($u=1$)}  \\
					\cmidrule(lr){5-7}\cmidrule(lr){8-10} \cmidrule(lr){11-13} \cmidrule(lr){14-16}
					\multicolumn{1}{c}{$\tau$} & \multicolumn{1}{c}{$\beta$} & \multicolumn{1}{c}{$n$} & \multicolumn{1}{c}{$r$}
					& \multicolumn{1}{c}{$\alpha$} & \multicolumn{1}{c}{$\lambda$} & \multicolumn{1}{c}{$\beta$} & \multicolumn{1}{c}{$\alpha$} & \multicolumn{1}{c}{$\lambda$} & \multicolumn{1}{c}{$\beta$} &\multicolumn{1}{c}{$\alpha$} & \multicolumn{1}{c}{$\lambda$} & \multicolumn{1}{c}{$\beta$} &\multicolumn{1}{c}{$\alpha$} & \multicolumn{1}{c}{$\lambda$} & \multicolumn{1}{c}{$\beta$} \\
					\midrule
					0.6& 0.35& 50& 30& 0.1503& -0.0692& 0.1388& -0.0787& -0.0309& 0.0521& -0.0784& -0.0308& 0.0519& -0.0804& -0.0334& 0.0563 \\
					& & & & (0.0691)&  (0.0307)& (0.0620)& (0.0319)& (0.0187)& (0.0436)& (0.0318)& (0.0187)& (0.0435)& (0.0322)& (0.0192)& (0.0444)  \\
					& & & 40& 0.1449& -0.0676& 0.1375& -0.0735& -0.0262& 0.0472& -0.0732& -0.0261& 0.0470&  -0.0746& -0.0276& 0.0495  \\
					& & & & (0.0665)& (0.0289)& (0.0605)& (0.0302)& (0.0179)& (0.0413)& (0.0301)& (0.0179)& (0.0412)&  (0.0308)& (0.0186)& (0.0423)   \\ [0.15 cm]
					& & 150& 60& 0.0863& -0.0358& 0.0617& -0.0448& -0.0208& 0.0283& -0.0446& -0.0206& 0.0284& -0.0487& -0.0226& 0.0296 \\
					& & & & (0.0455)& (0.0251)& (0.0514)& (0.0223)& (0.0159)& (0.0379)& (0.0221)&  (0.0159)&  (0.0379)& (0.0227)& (0.0169)& (0.0406)   \\
					& & & 120&  0.0693& -0.0321& 0.0512& -0.0327& -0.0174& 0.0241& -0.0325& -0.0173& 0.0241& -0.0359& -0.0182& 0.0267  \\
					& & & & (0.0339)& (0.0206)& (0.0321)& (0.0153)& (0.0132)& (0.0287)& (0.0151)& (0.0132)& (0.0286)& (0.0158)& (0.0146)&  (0.0295)  \\ [0.15 cm]
					& & 250& 100& 0.0468& -0.0258& 0.0436& -0.0231& 0.0115& -0.0179& -0.0229& 0.0114& -0.0178& -0.0251& 0.0126& -0.0193 \\
					& & & & (0.0286)& (0.0166)& (0.0259)& (0.0116)& (0.0107)& (0.0204)& (0.0115)& (0.0107)& (0.0203)&  (0.0120)& (0.0114)& (0.0212) \\
					& & & 200& 0.0371& -0.0194& 0.0368&  -0.0171& 0.0106& -0.0123& -0.0169& 0.0106& -0.0122& -0.0194& 0.0109& -0.0133   \\
					& & & & (0.0235)& (0.0120)& (0.0167)& (0.0109)& (0.0098)& (0.0175)& (0.0109)& (0.0098)& (0.0175)& (0.0118)& (0.0105)& (0.0194) \\
					\midrule
					& 0.70& 50& 30& 0.1006& -0.0450& 0.1335& -0.0788& 0.0325& 0.0534& -0.0785& 0.0324& 0.0532& -0.0845& 0.0332& 0.0564   \\
					& & & & (0.0541)& (0.0289)& (0.0568)& (0.0252)& (0.0194)& (0.0304)& (0.0251)& (0.0194)& (0.0303)& (0.0263)& (0.0203)& (0.0309)  \\
					& & & 40& 0.0833& -0.0360& 0.1259& -0.0682& 0.0306& 0.0508& -0.0679& 0.0306& 0.0507& -0.0732& 0.0311& 0.0511 \\
					& & & & (0.0523)& (0.0265)& (0.0535)& (0.0231)& (0.0171)& (0.0281)& (0.0230)& (0.0170)& (0.0280)& (0.0239)& (0.0189)& (0.0291) \\ [0.15 cm]
					& & 150& 60& 0.0654& -0.0332& 0.0654& -0.0384& 0.0237& 0.0467& -0.0382& 0.0236& 0.0465& -0.0390& 0.0246& 0.0478  \\
					& & & & (0.0459)&  (0.0209)& (0.0487)& (0.0216)& (0.0165)& (0.0233)& (0.0216)& (0.0165)& (0.0231)& (0.0221)& (0.0169)& (0.0247) \\
					& & & 120& 0.0577& -0.0314& 0.0530& -0.0297& 0.0216& -0.0517& -0.0295& 0.0215& -0.0516& -0.0306& 0.0227& -0.0551 \\
					& & & & (0.0326)& (0.0198)& (0.0415)& (0.0217)& (0.0136)& (0.0210)& (0.0216)& (0.0136)& (0.0209)& (0.0222)& (0.0139)& (0.0214) \\ [0.15 cm]
					& & 250& 100& 0.0452& -0.0308& 0.0467& -0.0214& 0.0165& -0.0388& -0.0213& 0.0164& -0.0387& -0.0226& 0.0173& -0.0392 \\
					& & & & (0.0284)& (0.0161)& (0.0283)& (0.0175)& (0.0112)& (0.0149)& (0.0174)& (0.0111)& (0.0148)& (0.0180)& (0.0119)& (0.0168) \\
					& & & 200& 0.0249& -0.0205& 0.0389& -0.0156& 0.0142& -0.0317& -0.0156& 0.0141& -0.0316& -0.0164& 0.0172& -0.0325  \\
					& & & & (0.0172)& (0.0123)& (0.0201)& (0.0106)& (0.0094)& (0.0121)& (0.0105)& (0.0094)& (0.0121)& (0.0109)& (0.0098)& (0.0129) \\ 
					\midrule 
					0.75& 0.35& 50& 30& 0.1123& -0.0507& 0.1333& -0.0594& 0.0293& 0.0427& -0.0591& 0.0292& 0.0428& -0.0603& 0.0307& 0.0433  \\
					& & & & (0.0487)& (0.0243)& (0.0582)& (0.0279)& (0.0177)& (0.0395)& (0.0279)& (0.0177)& (0.0395)& (0.0282)& (0.0181)& (0.0403) \\
					& & & 40& 0.1025& -0.0572& 0.1236& -0.0558& 0.0255& 0.0402& -0.0555& 0.0254& 0.0402& -0.0564& 0.0262& 0.0412 \\
					& & & & (0.0426)& (0.0229)& (0.0504)& (0.0248)& (0.0147)& (0.0350)& (0.0247)& (0.0146)& (0.0349)& (0.0152)& (0.0079)& (0.0358)\\ [0.15 cm]
					& & 150& 60& 0.0550& -0.0279& 0.0555& -0.0284& 0.0183& 0.0261& -0.0283& 0.0182& 0.0261& -0.0318& 0.0187& 0.0267  \\
					& & & & (0.0353)& (0.0195)& (0.0396)& (0.0190)& (0.0135)& (0.0251)& (0.0190)& (0.0135)& (0.0251)& (0.0192)& (0.0139)& (0.0257) \\
					& & & 120& 0.0453& -0.0243& 0.0487& -0.0243& 0.0149& 0.0197& -0.0242& 0.0148& 0.0197& -0.0252& 0.0157& 0.0204  \\
					& & & & (0.0324)& (0.0161)& (0.0317)& (0.0149)& (0.0123)& (0.0198)& (0.0149)& (0.0123)& (0.0197)& (0.0152)& (0.0126)& (0.0205)   \\ [0.15 cm]
					& & 250& 100& 0.0401& -0.0219& 0.0403& -0.0218& 0.0114& 0.0176 & -0.0217& 0.0114& 0.0176& -0.0224& 0.0120& 0.0184  \\
					& & & & (0.0292)& (0.0143)& (0.0281)& (0.0127)& (0.0107)& (0.0167)& (0.0127)& (0.0107)& (0.0167)& (0.0129)& (0.0111)& (0.0172) \\
					& & & 200& 0.0319& -0.0192& 0.0346& -0.0194& 0.0084& 0.0135& -0.0193& 0.0084& 0.0134& -0.0198& 0.0088& 0.0138  \\
					& & & & (0.0247)& (0.0121)& (0.0233)& (0.0107)& (0.0092)& (0.0144)& (0.0107)& (0.0092)& (0.0144)& (0.0110)& (0.0095)& (0.0148) \\
					\midrule
					& 0.70& 50& 30& 0.0710& -0.0473& 0.1160& -0.0481& 0.0249& -0.0388& -0.0478& 0.0247& -0.0387& -0.0535& 0.0263& -0.0393  \\
					& & & & (0.0324)& (0.0211)& (0.0553)& (0.0258)& (0.0135)& (0.0217)& (0.0257)& (0.0135)& (0.0217)& (0.0263)& (0.0139)& (0.0228)  \\
					& & & 40& 0.0671& -0.0430& 0.1051& -0.0452& 0.0214& -0.0350& -0.0450& 0.0213& -0.0348& -0.0502& 0.0225& -0.0358  \\
					& & & & (0.0308)& (0.0204)& (0.0531)& (0.0239)& (0.0119)& (0.0205)& (0.0238)& (0.0119)& (0.0204)& (0.0245)& (0.0124)& (0.0215) \\ [0.15 cm]
					& & 150& 60& 0.0523& -0.0248& 0.0529& -0.0268& 0.0178& -0.0228& -0.0267& 0.0178& -0.0227& -0.0282& 0.0183& -0.0236 \\
					& & & & (0.0267)& (0.0171)& (0.0341)& (0.0176)& (0.0104)& (0.0182)& (0.0175)& (0.0104)& (0.0181)& (0.0181)& (0.0110)& (0.0189)  \\
					& & & 120& 0.0438& -0.0219& 0.0462& -0.0250& 0.0145& -0.0189& -0.0249& 0.0145& -0.0188& -0.0258& 0.0150& -0.0193  \\
					& & & & (0.0241)& (0.0147)& (0.0305)& (0.0142)& (0.0098)& (0.0161)& (0.0141)& (0.0098)& (0.0160)& (0.0146)& (0.0102)& (0.0166) \\ [0.15 cm]
					& & 250& 100& 0.0343& -0.0183& 0.0363& -0.0195& 0.0111& -0.0151& -0.0194& 0.0111& -0.0150& -0.0217& 0.0116& -0.0159 \\
					& & & & (0.0190)& (0.0125)& (0.0263)& (0.0108)& (0.0089)& (0.0154)& (0.0107)& (0.0089)& (0.0154)& (0.0111)& (0.0092)& (0.0157) \\
					& & & 200& 0.0236& -0.0165& 0.0335& -0.0155& 0.0105& -0.0127& -0.0154& 0.0105& -0.0126& -0.0159& 0.0111& -0.0131 \\
					& & & & (0.0163)& (0.0101)& (0.0192)& (0.0092)& (0.0084)& (0.0118)& (0.0092)& (0.0084)& (0.0117)& (0.0096)& (0.0088)& (0.0125) \\
					\midrule
			\end{tabular}}
		\end{center}
		\vspace{-0.5cm}
	\end{table}

	\begin{table}[htbp!]
		\begin{center}
			\caption{AWs and CPs (in parantheses) of interval estimates for  the  Gumbel Type-II  baseline  lifetime  distribution  with  scale parameter $\lambda=0.75$ and shape parameter $\alpha=1$ for different choices of $\beta$ and $\tau$ under simple SSLT. }
			\label{T2}
			\tabcolsep 7pt
			\small
			\scalebox{0.63}{
				\begin{tabular}{*{16}c*{15}{r@{}l}}
					\toprule
					\multicolumn{4}{c}{} &
					\multicolumn{3}{c}{ACI} & \multicolumn{3}{c}{Boot-p} & \multicolumn{3}{c}{Boot-t}&\multicolumn{3}{c}{HPD}  \\
					\cmidrule(lr){5-7}\cmidrule(lr){8-10} \cmidrule(lr){11-13} \cmidrule(lr){14-16}
					\multicolumn{1}{c}{$\tau$} & \multicolumn{1}{c}{$\beta$} & \multicolumn{1}{c}{$n$} & \multicolumn{1}{c}{$r$}
					& \multicolumn{1}{c}{$\alpha$} & \multicolumn{1}{c}{$\lambda$} & \multicolumn{1}{c}{$\beta$} & \multicolumn{1}{c}{$\alpha$} & \multicolumn{1}{c}{$\lambda$} & \multicolumn{1}{c}{$\beta$} &\multicolumn{1}{c}{$\alpha$} & \multicolumn{1}{c}{$\lambda$} & \multicolumn{1}{c}{$\beta$} &\multicolumn{1}{c}{$\alpha$} & \multicolumn{1}{c}{$\lambda$} & \multicolumn{1}{c}{$\beta$} \\
					\midrule
					0.6& 0.35& 50& 30& 0.8460& 0.6748& 0.4534& 1.5960& 0.8149& 0.5938& 1.7956& 1.1379& 0.8718&  0.3684& 0.2291&  0.2227 \\
					& & & & (0.9571)& (0.9060)& (0.9461)& (0.9799)& (0.9583)& (0.9543)& (0.9395)& (0.9146)& (0.9424)& (0.9627)& (0.9419)& (0.9526)  \\
					& & & 40& 0.8169& 0.6569& 0.4416& 1.6294& 0.7131& 0.5355& 1.8001& 1.1156& 0.8932& 0.3647& 0.2259&  0.2204  \\
					& & & & (0.9460)& (0.8952)& (0.9389)& (0.9695)& (0.9517)& (0.9529)& (0.9289)& (0.8894)& (0.9385)&  (0.9544)& (0.9253)& (0.9478)  \\ [0.15 cm] 
					& & 150& 60& 0.4672& 0.4145& 0.4300& 0.6713& 0.5652& 0.3681& 1.6115& 1.1636& 0.7002& 0.3019& 0.2075& 0.1929  \\
					& & & &  (0.9583)& (0.9282)& (0.9344)& (0.9867)& (0.9477)& (0.9473)& (0.9656)& (0.9363)& (0.9274)& (0.9635)& (0.9396)& (0.9395)  \\
					& & & 120& 0.4364& 0.3992& 0.4019& 0.6977& 0.4201& 0.2078& 1.6198& 1.1651& 0.6652&  0.2759& 0.2083& 0.1721  \\
					& & & & (0.9552)& (0.9395)& (0.9472)& (0.9850)& (0.9567)& (0.9793)&  (0.9486)& (0.9262)& (0.9360)& (0.9769)& (0.9410)& (0.9580)  \\ [0.15 cm]
					& & 250& 100& 0.3615& 0.3231& 0.3797& 0.4937&  0.4394& 0.2245& 1.5871& 1.1496& 0.6379& 0.2630& 0.1929& 0.1634 \\
					& & & & (0.9595)& (0.9460)& (0.9536)& (0.9772)& (0.9629)& (0.9825)& (0.9083)& (0.9286)& (0.9119)& (0.9629)& (0.9603)& (0.9742)   \\
					& & & 200& 0.3347& 0.3088& 0.3022&  0.5287&0.3351 & 0.2461& 1.6058&1.1389 & 0.6206& 0.2371& 0.1909& 0.1475 \\
					& & & & (0.9496)& (0.9485)& (0.9588)& (0.9859)& (0.9608)& (0.9899)& (0.9397)& (0.9509)& (0.9427)& (0.9605)& (0.9534)& (0.9793) \\ 
					\midrule
					& 0.70& 50& 30& 0.8191& 0.6825& 0.4933& 1.2686& 0.7207& 0.7336& 1.7295& 1.1337& 1.2294& 0.3711& 0.2427& 0.2494   \\
					& & & & (0.9784)& (0.9324)& (0.9577)&  (0.9879)& (0.9685)& (0.9775)& (0.9762)& (0.9174)& (0.9474)& (0.9830)& (0.9551)& (0.9678) \\
					& & & 40& 0.7718& 0.6773& 0.4796& 1.3073& 0.8221& 0.6678& 1.6823& 1.1829& 1.2592& 0.3486& 0.2375& 0.2507 \\
					& & & & (0.9487)& (0.9493)& (0.9381)& (0.9796)& (0.9598)& (0.9798)&  (0.9484)& (0.9514)& (0.9456)& (0.9563)& (0.9602)& (0.9459) \\ [0.15 cm]
					& & 150& 60& 0.4678& 0.4172& 0.4558& 0.6533& 0.5504& 0.6218& 1.6178& 1.1628& 1.2682& 0.2983& 0.2162& 0.2800 \\
					& & & & (0.9469)& (0.9485)& (0.9635)& (0.9756)& (0.9645)& (0.9884)& (0.9497)& (0.9483)& (0.9543)& (0.9662)& (0.9526)& (0.9451)  \\
					& & & 120& 0.4355& 0.3983& 0.3807& 0.6921& 0.4174& 0.3942& 1.6244& 1.1579& 1.2605& 0.2674& 0.2082& 0.2687  \\
					& & & & (0.9586)& (0.9433)& (0.9472)& (0.9923)& (0.9782)& (0.9836)& (0.9498)& (0.9462)& (0.9414)& (0.9658)& (0.9524)& (0.9625) \\ [0.15 cm]
					& & 250& 100& 0.3596& 0.3218& 0.3234& 0.4919& 0.4382& 0.4270& 1.5843& 1.1482& 1.2203& 0.2762& 0.2047& 0.2993  \\
					& & & & (0.9521)& (0.9508)& (0.9464)& (0.9832)& (0.9625)& (0.9805)& (0.9411)& (0.9427)& (0.9281)& (0.9616)& (0.9545)& (0.9515) \\
					& & & 200& 0.3341& 0.3099& 0.3199& 0.5261& 0.3345& 0.3872& 1.5894& 1.1598& 1.2190& 0.2563& 0.2073& 0.2888 \\
					& & & & (0.9467)& (0.9416)& (0.9574)& (0.9796)& (0.9601)& (0.9853)& (0.9498)& (0.9459)& (0.9401)& (0.9517)& (0.9542)& (0.9548)  \\
					\midrule
					0.75& 0.35& 50& 30& 0.7355& 0.6006& 0.4793& 1.3131& 0.6987& 0.6078& 1.7811& 1.1371& 0.9333& 0.3666& 0.2287& 0.2176 \\
					& & & & (0.9527)& (0.9419)& (0.9577)& (0.9815)& (0.9725)& (0.9798)& (0.9443)& (0.9424)& (0.9591)& (0.9665)& (0.9527)& (0.9689)  \\
					& & & 40& 0.7097& 0.5988& 0.4606& 1.3036& 0.6099& 0.5271& 1.7371& 1.1670& 0.8620& 0.3608& 0.2292& 0.2162 \\
					& & & & (0.9614)& (0.9522)& (0.9572)& (0.9899)& (0.9688)& (0.9799)& (0.9600)& (0.9475)& (0.9667)& (0.9637)& (0.9623)& (0.9697)  \\ [0.15 cm]
					& & 150& 60& 0.4098& 0.3616& 0.4380& 0.6778& 0.4669& 0.4572& 1.6333& 1.1461& 0.7380& 0.2845& 0.2026& 0.2048  \\
					& & & & (0.9425)& (0.9370)& (0.9494)& (0.9789)& (0.9657)& (0.9799)& (0.9594& (0.9472)& (0.9612)& (0.9717)& (0.9533)& (0.9701) \\ 
					& & & 120& 0.3959& 0.3516& 0.3865& 0.6377& 0.3694& 0.4324& 1.6370& 1.1357& 0.6652& 0.2641& 0.2014& 0.1774 \\
					& & & & (0.9438)& (0.9326)& (0.9466)& (0.9854)& (0.9732)& (0.9785)& (0.9499)& (0.9592)& (0.9426)& (0.9820)& (0.9520)& (0.9615) \\ [0.15 cm]
					& & 250& 100& 0.3123& 0.2825& 0.3689& 0.4933& 0.3423& 0.4174& 1.5624& 1.1594& 0.5998& 0.2368& 0.1825& 0.1726  \\
					& & & & (0.9364)& (0.9463)& (0.9614)& (0.9698)& (0.9529)& (0.9781)& (0.9499)& (0.9488)& (0.9605)& (0.9585)& (0.9508)& (0.9708)  \\
					& & & 200& 0.3017& 0.2757& 0.2892& 0.4907& 0.2832& 0.3646& 1.5871& 1.1494& 0.6183& 0.2289& 0.1830& 0.1651  \\
					& & & & (0.9425)& (0.9438)& (0.9552)& (0.9751)& (0.9656)& (0.9785)& (0.9498)& (0.9548)& (0.9602)& (0.9676)& (0.9520)& (0.9752)  \\
					\midrule
					& 0.70& 50& 30& 0.7158& 0.6178& 0.4900& 1.1718& 0.6416& 0.7814& 1.6753& 1.1838& 1.2681& 0.3612& 0.2384& 0.2631  \\
					& & & & (0.9374)& (0.9452)& (0.9592)& (0.9738)& (0.9678)& (0.9896)& (0.9384)& (0.9481)& (0.9441)& (0.9605)& (0.9507)& (0.9660)  \\
					& & & 40& 0.6945& 0.6072& 0.4779& 1.1588& 0.5806& 0.6380& 1.6403& 1.1697& 1.3338& 0.3471& 0.2378& 0.2562  \\
					& & & & (0.9457)& (0.9411)& (0.9477)& (0.9875)& (0.9689)& (0.9792)& (0.9486)& (0.9462)& (0.9672)& (0.9597)& (0.9589)& (0.9578)  \\ [0.15 cm]
					& & 150& 60& 0.4070& 0.3602& 0.3968& 0.6295& 0.4241& 0.4396& 1.6007& 1.1449& 1.2600& 0.2795& 0.2047& 0.2322  \\
					& & & & (0.9562)& (0.9610)& (0.9473)& (0.9875)& (0.9764)& (0.9699)& (0.9499)& (0.9569)& (0.9289)& (0.9687)& (0.9648)& (0.9609)  \\
					& & & 120& 0.3919& 0.3540& 0.3756& 0.6318& 0.3545& 0.4129& 1.6154& 1.1536& 1.2624& 0.2644& 0.2054& 0.2234 \\
					& & & & (0.9462)& (0.9475)& (0.9483)& (0.9755)& (0.9762)& (0.9782)& (0.9492)& (0.9511)& (0.9505)& (0.9676)& (0.9585)& (0.9631)  \\ [0.15 cm]
					& & 250& 100& 0.3120& 0.2835& 0.3871& 0.4873& 0.3357& 0.4212& 1.5703& 1.1562& 1.1857& 0.2413& 0.1873& 0.2832 \\
					& & & & (0.9552)& (0.9420)& (0.9458)& (0.9766)& (0.9639)& (0.9780)& (0.9521)& (0.9494)& (0.9404)& (0.9699)& (0.9549)& (0.9629) \\
					& & & 200& 0.3019& 0.2777& 0.3682& 0.4912& 0.2819& 0.3847& 1.5572& 1.1503& 1.1692& 0.2339& 0.1841& 0.2737 \\
					& & & & (0.9560)& (0.9448)& (0.9427)& (0.9847)& (0.9667)& (0.9749)& (0.9523)& (0.9566)& (0.9438)& (0.9684)& (0.9610)& (0.9623)  \\
					\midrule
			\end{tabular}}
		\end{center}
		\vspace{-0.5cm}
	\end{table}

	\begin{table}[htbp!]
		\begin{center}
			\caption{ABs and MSEs (in parantheses) of MLEs and Bayes estimates for  the  Gumbel Type-II  baseline  lifetime  distribution  with  scale parameter $\lambda=0.75$ and shape parameter $\alpha=1.5$ for different choices of $\beta$ and $\tau$ under simple SSLT. }
			\label{T3}
			\tabcolsep 7pt
			\small
			\scalebox{0.63}{
				\begin{tabular}{*{16}c*{15}{r@{}l}}
					\toprule
					\multicolumn{4}{c}{} &
					\multicolumn{3}{c}{MLE} & \multicolumn{3}{c}{SEL} & \multicolumn{3}{c}{LL~($u=-0.05$)}&\multicolumn{3}{c}{LL~($u=1$)}  \\
					\cmidrule(lr){5-7}\cmidrule(lr){8-10} \cmidrule(lr){11-13} \cmidrule(lr){14-16}
					\multicolumn{1}{c}{$\tau$} & \multicolumn{1}{c}{$\beta$} & \multicolumn{1}{c}{$n$} & \multicolumn{1}{c}{$r$}
					& \multicolumn{1}{c}{$\alpha$} & \multicolumn{1}{c}{$\lambda$} & \multicolumn{1}{c}{$\beta$} & \multicolumn{1}{c}{$\alpha$} & \multicolumn{1}{c}{$\lambda$} & \multicolumn{1}{c}{$\beta$} &\multicolumn{1}{c}{$\alpha$} & \multicolumn{1}{c}{$\lambda$} & \multicolumn{1}{c}{$\beta$} &\multicolumn{1}{c}{$\alpha$} & \multicolumn{1}{c}{$\lambda$} & \multicolumn{1}{c}{$\beta$} \\
					\midrule
					0.6& 0.35& 50& 30& 0.3096& -0.1097& 0.1424& -0.1003& -0.0592& 0.0478& -0.0998& -0.0591& 0.0476& -0.1090& -0.0617& 0.0485  \\
					& & & & (0.2944)& ( 0.0606)& (0.0715)& (0.0644)& (0.0387)& (0.0365)& (0.0642)& (0.0386)& (0.0364)& (0.0665)& (0.0389)& (0.0378)  \\
					& & & 40& 0.2836& -0.0962& 0.1380& -0.0851& -0.0542& 0.0379& -0.0847& -0.0540& 0.0377& -0.0861& -0.0551& 0.0389 \\ 
					& & & & (0.2631)& (0.0556)& (0.0655)& (0.0558)& (0.0352)& (0.0337)& (0.0557)& (0.0351)& (0.0336)& (0.0575)& (0.0359)& (0.0372) \\ [0.15 cm]
					& & 150& 60& 0.0935& -0.0622& 0.0736& -0.0623& -0.0362& 0.0339& -0.0620& -0.0361& 0.0337& -0.0638& -0.0372& 0.0351 \\
					& & & & (0.0799)& (0.0478)& (0.0552)& (0.0470)& (0.0326)& (0.0308)& (0.0469)& (0.0325)& (0.0306)& (0.0478)& (0.0335)& (0.0315) \\
					& & & 120& 0.0850& -0.0572& 0.0677& -0.0605& -0.0295& 0.0318& -0.0602& -0.0294& 0.0318& -0.0658& -0.0304& 0.0326 \\
					& & & & (0.0677)& (0.0381)& (0.0441)& (0.0417)& (0.0287)& (0.0255)& (0.0416)& (0.0286)& (0.0255)& (0.0425)& (0.0295)& (0.0264) \\ [0.15 cm]
					& & 250& 100& 0.0637& -0.0245& 0.0364& -0.0467& -0.0108& 0.0255& -0.0465& -0.0108& 0.0255& -0.0519& -0.0124& 0.0269 \\
					& & & & (0.0356)& (0.0116)& (0.0071)& (0.0194)& (0.0046)& (0.0051)& (0.0193)& (0.0046)& (0.0051)& (0.0212)& (0.0048)& (0.0054) \\
					& & & 200& 0.0532& -0.0203& 0.0255& -0.0384& -0.0091& 0.0229& -0.0342& -0.0090& 0.0229& -0.0391& -0.0107& 0.0235  \\
					& & & & (0.0269)& (0.0094)& (0.0061)& (0.0172)& (0.0076)& (0.0050)& (0.0171)& (0.0076)& (0.0050)& (0.0199)& (0.0078)& (0.0053) \\
					\midrule
					& 0.70& 50& 30& 0.2145& -0.0728& 0.1475& -0.0331& 0.0177& -0.0225& -0.0327& 0.0178& -0.0225& -0.0419& 0.0185& -0.0231  \\
					& & & & (0.1417)& (0.0337)& (0.0588)& (0.0301)& (0.0090)& (0.0169)& (0.0300)& (0.0090)& (0.0168)& (0.0306)& (0.0092)& (0.0177)  \\
					& & & 40& 0.1534& -0.0431& 0.1215& -0.0469& 0.0196& -0.0190& -0.0465& 0.0196& -0.0190& -0.0551& 0.0205& -0.0198 \\
					& & & & (0.1083)& (0.0346)& (0.0559)& (0.0265)& (0.0086)& (0.0162)& (0.026)& (0.0085)& (0.0161)& (0.0273)& (0.0089)& (0.0168)  \\ [0.15 cm]
					& & 150& 60& 0.0708& -0.0225& 0.0628& -0.0396& 0.0168& -0.0184& -0.039& 0.0168& -0.0184& -0.0452& 0.0175& -0.0191  \\
					& & & & (0.0504)& (0.0160)& (0.0330)& (0.0182)& (0.0071)& (0.0139)& (0.0182)& (0.0071)& (0.0139)& (0.0187)& (0.0077)& (0.0145) \\
					& & & 120& 0.0660& -0.0207& 0.0510& -0.0329& 0.0135& -0.0167& -0.0327& 0.135& -0.0166& -0.0376& 0.0166& -0.0188 \\
					& & & & (0.0392)& (0.0138)& (0.0319)& (0.0143)& (0.0058)& (0.0116)& (0.0143)& (0.0058)& (0.0116)& (0.0149)& (0.0061)& (0.0121) \\ [0.15 cm]
					& & 250& 100& 0.0486& -0.0152& 0.0388& -0.0220& 0.0124& -0.0149& -0.0218& 0.0124& -0.0148& -0.0268& 0.0145& -0.0163 \\
					& & & & (0.0301)& (0.0106)& (0.0233)& (0.0135)& (0.0055)& (0.0122)& (0.0135)& (0.0055)& (0.0122)& (0.0139)& (0.0060)& (0.0127) \\
					& & & 200& 0.0472& -0.0141& 0.0301& -0.0179& 0.0111& -0.0125& -0.0178& 0.0111& -0.0125& -0.0205& 0.0119& -0.0143  \\
					& & & & (0.0261)& (0.0101)& (0.0218)& (0.0119)& (0.0053)& (0.0103)& (0.0118)& (0.0053)& (0.0103)& (0.0124)& (0.0055)& (0.0108) \\
					\midrule
					0.75& 0.35& 50& 30& 0.1812& -0.0509& 0.1115& 0.0133& -0.0075& 0.0407& 0.0138& -0.0074& 0.0407& 0.0147& -0.0101& 0.0417  \\
					& & & & (0.1398)& (0.0311)& (0.0477)& (0.0285)& (0.0081)& (0.0068)& (0.0286)& (0.0081)& (0.0068)& (0.0297)& (0.0082)& (0.0070) \\
					& & & 40& 0.1789& -0.0455& 0.1022& 0.0264& -0.0096& 0.0326& 0.0267& -0.0096& 0.0327& 0.0281& -0.0121& 0.0338  \\
					& & & & (0.1303)& (0.0304)& (0.0431)& (0.0234)& (0.0078)& (0.0054)& (0.0234)& (0.0078)& (0.0054)& (0.0241)& (0.0081)& (0.0056)  \\ [0.15 cm]
					& & 150& 60& 0.0750& -0.0257& 0.0410& 0.0317& -0.0120& 0.0283& 0.0320& -0.0120& 0.0282& 0.0342& -0.0139& 0.0295    \\
					& & & & (0.0337)& (0.0110)& (0.0119)& (0.0162)& (0.0057)& (0.0068)& (0.0162)& (0.0057)& (0.0068)& (0.0176)& (0.0058)& (0.0070)   \\
					& & & 120& 0.0668& -0.0208& 0.0366& 0.0257& -0.0073& 0.0253& 0.0260& -0.0073& 0.0254& 0.0277& -0.0091& 0.0260 \\
					& & & & (0.0319)& (0.0094)& (0.0093)& (0.0149)& (0.0054)& (0.0050)& (0.0150)& (0.0054)& (0.0050)& (0.0155)& (0.0056)& (0.0051) \\ [0.15 cm]
					& & 250& 100& 0.0298& -0.0094& 0.0384& 0.0130& -0.0074& 0.0340& 0.0132& -0.0074& 0.0341& 0.0146& -0.0082& 0.0360 \\
					& & & & (0.0189)& (0.0061)& (0.0141)& (0.0134)& (0.0048)& (0.0098)& (0.0134)& (0.0048)& (0.0098)& (0.0137)& (0.0049)& (0.0101)  \\
					& & & 200& 0.0232& -0.0082& 0.0208& 0.0110& -0.0059& 0.0190& 0.0110& -0.0059& 0.0189& 0.0121& -0.0074& 0.0204 \\
					& & & & (0.0173)& (0.0058)& (0.0109)& (0.0122)& (0.0043)& (0.0087)& (0.0122)& (0.0043)& (0.0087)& (0.0130)& (0.0044)& (0.0096)  \\
					\midrule
					& 0.70& 50& 30& 0.1287& -0.0328& 0.1001& -0.0359& 0.0147& -0.0341& -0.0356& 0.0145& -0.0335& -0.0443& 0.0154& -0.0362  \\
					& & & & (0.0925)& (0.0251)& (0.0518)& (0.0242)& (0.0161)& (0.0191)& (0.0242)& (0.0161)& (0.0190)& (0.0251)& (0.0183)& (0.0200) \\
					& & & 40& 0.1156& -0.0315& 0.0946& -0.0323& 0.0131& -0.0286& -0.0321& 0.0131& -0.0285& -0.0372& 0.0135& -0.0301  \\
					& & & & (0.0885)& (0.0237)& (0.0502)& (0.0235)& (0.0101)& (0.0178)& (0.0234)& (0.0101)& (0.0178)& (0.0241)& (0.0108)& (0.0183) \\ [0.15 cm]
					& & 150& 60& 0.0686& -0.0235& 0.0402& -0.0305& 0.0118& -0.0273& -0.0302& 0.0118& -0.0272& -0.0329& 0.0129& -0.0286  \\
					& & & & (0.0598)&(0.0194) & (0.0284)& (0.0195)& (0.0085)& (0.0166)& (0.0195)& (0.0085)& (0.0165)& (0.0199)& (0.0092)& (0.0173) \\
					& & & 120& 0.0568& -0.0192& 0.0360& -0.0283& 0.0107& -0.0235& -0.0281& 0.0107& -0.0234& -0.0295& 0.0118& -0.0268  \\
					& & & & (0.0395)& (0.0176)& (0.0251)& (0.0136)& (0.0067)& (0.0152)& (0.0136)& (0.0067)& (0.0152)& (0.0141)& (0.0071)& (0.0169)  \\ [0.15 cm]
					& & 250& 100& 0.0395& -0.0143& 0.0331& -0.0124& 0.0101& -0.0204& -0.0122& 0.0101& -0.0202& -0.0138& 0.0111& -0.0221    \\
					& & & & (0.0210)& (0.0154)& (0.0231)& (0.0123)& (0.0054)& (0.0139)& (0.0123)& (0.0054)& (0.0139)& (0.0135)& (0.0056)& (0.0153)   \\
					& & & 200& 0.0335& -0.0128& 0.0279& -0.0106& 0.0074& -0.0158& -0.0105& 0.0074& -0.0156& -0.0124& 0.0085& -0.0187  \\
					& & & & (0.0171)& (0.0127)& (0.0176)& (0.0115)& (0.0050)& (0.0115)& (0.0115)& (0.0050)& (0.0115)& (0.0121)& (0.0051)& (0.0120) \\		
					\midrule
			\end{tabular}}
		\end{center}
		\vspace{-0.5cm}
	\end{table}

	\begin{table}[htbp!]
		\begin{center}
			\caption{AWs and CPs (in parantheses) of interval estimates for  the  Gumbel Type-II  baseline  lifetime  distribution  with  scale parameter $\lambda=0.75$ and shape parameter $\alpha=1.5$ for different choices of $\beta$ and $\tau$ under simple SSLT. }
			\label{T4}
			\tabcolsep 7pt
			\small
			\scalebox{0.63}{
				\begin{tabular}{*{16}c*{15}{r@{}l}}
					\toprule
					\multicolumn{4}{c}{} &
					\multicolumn{3}{c}{ACI} & \multicolumn{3}{c}{Boot-p} & \multicolumn{3}{c}{Boot-t}&\multicolumn{3}{c}{HPD}  \\
					\cmidrule(lr){5-7}\cmidrule(lr){8-10} \cmidrule(lr){11-13} \cmidrule(lr){14-16}
					\multicolumn{1}{c}{$\tau$} & \multicolumn{1}{c}{$\beta$} & \multicolumn{1}{c}{$n$} & \multicolumn{1}{c}{$r$}
					& \multicolumn{1}{c}{$\alpha$} & \multicolumn{1}{c}{$\lambda$} & \multicolumn{1}{c}{$\beta$} & \multicolumn{1}{c}{$\alpha$} & \multicolumn{1}{c}{$\lambda$} & \multicolumn{1}{c}{$\beta$} &\multicolumn{1}{c}{$\alpha$} & \multicolumn{1}{c}{$\lambda$} & \multicolumn{1}{c}{$\beta$} &\multicolumn{1}{c}{$\alpha$} & \multicolumn{1}{c}{$\lambda$} & \multicolumn{1}{c}{$\beta$} \\
					\midrule
					0.6& 0.35& 50& 30& 1.6347& 0.8261& 0.8242& 2.3538& 1.3013& 0.5756& 3.6895& 1.8203& 1.1368& 0.4418& 0.2312& 0.1366 \\
					& & & & (0.9435)& (0.9385)& (0.9481)& (0.9616)& (0.9576)& (0.9726)& (0.9408)& (0.9338)& (0.9565)& (0.9581)& (0.9506)& (0.9685) \\
					& & & 40& 1.5664& 0.8208& 0.8130& 2.1842& 1.1994& 0.5254& 3.3956& 2.9568& 1.1120& 0.4356& 0.2294& 0.1381  \\
					& & & & (0.9539)& (0.9551)& (0.9566)& (0.9752)& (0.9721)& (0.9673)& (0.9488)& (0.9532)& (0.9483)& (0.9603)& (0.9691)& (0.9625) \\
					& & 150& 60& 0.8705& 0.5159& 0.4323& 1.2824& 0.9260& 0.4335& 2.4847& 1.1545& 0.6699& 0.3911& 0.2064& 0.1281 \\
					& & & & (0.9516)& (0.9241)& (0.9461)& (0.9732)& (0.9548)& (0.9686)& (0.9474)& (0.9224)& (0.9589)& (0.9684)& (0.9434)& (0.9623)  \\
					& & & 120& 0.8078& 0.4907& 0.3981& 1.4582& 0.7172& 0.3989& 2.5399& 1.1395& 0.6817& 0.3526& 0.2094& 0.1221  \\
					& & & & (0.9438)& (0.9506)& (0.9463)& (0.9647)& (0.9805)& (0.9679)& (0.9381)& (0.9615)& (0.9568)& (0.9595)& (0.9763)& (0.9632) \\ [0.15 cm]
					& & 250& 100& 0.6653& 0.4017& 0.3206& 0.8556& 0.7086& 0.2851& 2.4474& 1.1547& 0.6113& 0.3538& 0.1881& 0.1162 \\
					& & & & (0.9435)& (0.93416)& (0.9462)& (0.9649)& (0.9518)& (0.9699)& (0.9401)& (0.9373)& (0.9511)& (0.9572)& (0.9431)& (0.9623) \\
					& & & 200& 0.6172& 0.3873& 0.2975& 0.9969& 0.5544& 0.2653& 2.4298& 1.1541& 0.6116& 0.3255& 0.1921& 0.1150 \\
					& & & & (0.9515)& (0.9437)& (0.9462)& (0.9742)& (0.9608)& (0.9699)& (0.9491)& (0.9489)& (0.9502)& (0.9609)& (0.9584)& (0.9583) \\
					\midrule
					& 0.70& 50& 30& 1.5655& 0.8514& 1.3980& 1.2818& 0.6418& 0.5403& 2.7711& 1.1407& & 0.4443& 0.2408& 0.2516 \\
					& & & & (0.9492)& (0.9458)& (0.9487)& (0.9686)& (0.9731)& (0.9757)& (0.9471)& (0.9580)& (0.9493)& (0.9534)& (0.9612)& (0.9691)  \\
					& & & 40& 1.4489& 0.8495& 1.3750& 1.1231& 0.6415& 0.5551& 2.7080& 1.1893& 1.1905& 0.4306& 0.2416& 0.2485 \\
					& & & & (0.9482)& (0.9580)& (0.9542)& (0.9819)& (0.9743)& (0.9777)& (0.9479)& (0.9485)& (0.9494)& (0.9643)& (0.9623)& (0.9603) \\ [0.15 cm]
					& & 150& 60& 0.8611& 0.5192& 0.8304& 0.8117& 0.4619& 0.5129& 2.4752& 1.1774& 1.0908& 0.3668& 0.2144& 0.2324 \\
					& & & & (0.9561)& (0.9437)& (0.9479)& (0.9716)& (0.9631)& (0.9677)& (0.9493)& (0.9495)& (0.9494)& (0.9642)& (0.9527)& (0.9532) \\
					& & & 120& 0.8038& 0.4953& 0.7778& 0.6871& 0.4210& 0.4862& 2.4626& 1.1478& 1.1094& 0.3380& 0.2091& 0.2222  \\
					& & & & (0.9382)& (0.9457)& (0.9572)& (0.9611)& (0.9647)& (0.9835)& (0.9497)& (0.9442)& (0.9595)& (0.9593)& (0.9549)& (0.9736)  \\ [0.15 cm]
					& & 250& 100& 0.6615& 0.4055& 0.6275& 0.6347& 0.3745& 0.4677& 2.4202& 1.1619& 1.0871& 0.3391& 0.2038& 0.2215 \\
					& & & & (0.9535)& (0.9364)& (0.9569)& (0.9710)& (0.9538)& (0.9762)& (0.9492)& (0.9435)& (0.9599)& (0.9685)& (0.9529)& (0.9728) \\ 
					& & & 200& 0.6155& 0.3886& 0.5946& 0.5542& 0.3463& 0.4435& 2.4039& 1.1527& 1.1148& 0.2985& 0.1955& 0.2112  \\
					& & & & (0.9456)& (0.9433)& (0.9473)& (0.9609)& (0.9626)& (0.9799)& (0.9496)& (0.9496)& (0.9596)& (0.9573)& (0.9545)& (0.9647) \\
					\midrule
					0.75& 0.35& 50& 30& 1.2109& 0.6579& 0.7493& 1.3536& 0.6052& 0.6980& 2.7070& 1.1750& 0.8073& 0.4417& 0.2341& 0.2137  \\
					& & & & (0.9413)& (0.9460)& (0.9371)& (0.9678)& (0.9789)& (0.9673)& (0.9395)& (0.9485)& (0.9412)& (0.9548)& (0.9533)& (0.9644)   \\
					& & & 40& 1.2016& 0.6587& 0.6855& 1.3167& 0.6003& 0.6600& 2.756& 1.1615& 0.7987& 0.4281& 0.2337& 0.2033 \\
					& & & & (0.9459)& (0.9487)& (0.9369)& (0.9682)& (0.9756)& (0.9642)& (0.9455)& (0.9483)& (0.9498)& (0.9643)& (0.9603)& (0.9517) \\ [0.15 cm]
					& & 150& 60& 0.6671& 0.3930& 0.3956& 0.6930& 0.3826& 0.4505& 2.4542& 1.1540& 0.6513& 0.3617& 0.2034& 0.1948   \\
					& & & & (0.9486)& (0.9322)& (0.9523)& (0.9659)& (0.9619)& (0.9742)& (0.9497)& (0.9395)& (0.9428)& (0.9564)& (0.9504)& (0.9627)   \\
					& & & 120&  0.6537& 0.3870& 0.3489& 0.6812& 0.3760& 0.3788& 2.4388& 1.1510& 0.6377& 0.3480& 0.2017& 0.1786   \\
					& & & & (0.9529)& (0.9356)& (0.9457)& (0.9686)& (0.9635)& (0.9603)& (0.9492)& (0.9491)& (0.9439)& (0.9675)& (0.9540)& (0.9581)  \\ [0.15 cm]
					& & 250& 100& 0.5129& 0.3073& 0.4215& 0.6077& 0.3559& 0.5747& 2.2909& 1.1403& 0.6890& 0.3272& 0.1854& 0.2065 \\
					& & & & (0.9543)& (0.9515)& (0.9466)& (0.9735)& (0.9743)& (0.9683)& (0.9489)& (0.9585)& (0.9509)& (0.9672)& (0.9641)& (0.9594) \\
					& & & 200& 0.4993& 0.3034& 0.2621& 0.5105& 0.2970& 0.2750& 2.3674& 1.1548& 0.5971& 0.3097& 0.1824& 0.1607 \\
					& & & & (0.9493)& (0.9436)& (0.9561)& (0.9629)& (0.9625)& (0.9719)& (0.9495)& (0.9496)& (0.9479)& (0.9562)& (0.9537)& (0.9618) \\
					\midrule
					& 0.70& 50& 30& 1.1796& 0.6641& 1.2519& 1.1490& 0.5880& 0.5810& 2.5727& 1.1875& 1.1396& 0.4367& 0.2387& 0.2649  \\
					& & & & (0.9467)& (0.9419)& (0.9574)& (0.9694)& (0.9639)& (0.9735)& (0.9473)& (0.9486)& (0.9488)& (0.9651)& (0.9517)& (0.9691) \\
					& & & 40& 1.1717& 0.6552& 1.2258& 1.0097& 0.5587& 0.5324& 2.6251& 1.1533& 1.1736& 0.4246& 0.2387& 0.2597 \\
					& & & & (0.9532)& (0.9419)& (0.9477)& (0.9748)& (0.9629)& (0.9676)& (0.9483)& (0.9493)& (0.9493)& (0.9645)& (0.9522)& (0.9581) \\ [0.15 cm]
					& & 150& 60& 0.6654& 0.3936& 0.7660& 0.6454& 0.3685& 0.5117& 2.4152& 1.1534& 1.0870& 0.3611& 0.2079& 0.2476   \\
					& & & & (0.9445)& (0.9395)& (0.9566)& (0.9626)& (0.9634)& (0.9860)& (0.9396)& (0.9497)& (0.9596)& (0.9576)& (0.9533)& (0.9798)  \\
					& & & 120& 0.6519& 0.3891& 0.6860& 0.5946& 0.3534& 0.4670& 2.4343& 1.1501& 1.1023& 0.3385& 0.2058& 0.2259  \\
					& & & & (0.9455)& (0.9430)& (0.9576)& (0.9618)& (0.9642)& (0.9755)& (0.9496)& (0.9468)& (0.9495)& (0.9571)& (0.9532)& (0.9685) \\ [0.15 cm]
					& & 250& 100& 0.5135& 0.3045& 0.6306& 0.6559& 0.3500& 0.4319& 2.2821& 1.1406& 1.0532& 0.3248& 0.1883& 0.2204   \\
					& & & & (0.9485)& (0.9452)& (0.9560)& (0.9625)& (0.9641)& (0.9839)& (0.9478)& (0.9487)& (0.9491)& (0.9586)& (0.9526)& (0.9612) \\
					& & & 200& 0.4995& 0.3037& 0.5211& 0.4744& 0.2861& 0.4248& 2.2736& 1.1323& 1.0260& 0.2931& 0.1855& 0.2156  \\
					& & & & (0.9524)& (0.9530)& (0.9478)& (0.9719)& (0.9646)& (0.9659)& (0.9499)& (0.9425)& (0.9495)& (0.9570)& (0.9539)& (0.9586) \\ 
					\midrule
			\end{tabular}}
		\end{center}
		\vspace{-0.5cm}
	\end{table}
	\section{Optimality Criteria}
	In reliability and survival analysis, an optimum censoring plan among chosen schemes is desired to get a sufficient amount of information about the unknown model parameters. However, comparing two (or more)
		different censoring plans has gained a lot of attention in past few years
		by several authors. For instance one can see Ka et al. \cite{ka2011optimal}, Guan and Tang \cite{guan2012optimal}, Singh et al. \cite{singh2015estimating}, Abd El-Raheem \cite{abd2018optimal}, Hakamipour \cite{hakamipour2021comparison} and Dutta and Kayal \cite{dutta2022inference}. Here, three commonly used criteria have been considered based on the variance-covariance matrix (VCM) of the observed Fisher information matrix corresponding to the MLEs of unknown parameters (see Table $5$). 
	    \subsection*{$A$-optimality}
		This first criterion is based on the trace of the first order approximation of the variance-covariance matrix (VCM) of the MLEs. The trace of the VCM equals to the sum of the diagonal elements of $I^{-1}(\widehat{\Theta})$. This $A$- optimality criterion provides an overall measure of the average variance of the estimates under MLE. The $A$- optimality criterion is defined as $minimizing$ $trace$ $I^{-1}(\widehat{\Theta})$. 
		\subsection*{$D$-optimality}
		This second criterion is based on maximizing the determinant of the observed Fisher information matrix which is equivalent to minimize the determinant of VCM. We know that, the joint confidence region of $\Theta$ is proportional to $|I^{-1}(\widehat{\Theta})|^{1/2}$ under some fixed level of confidence. So smaller value of $|I^{-1}(\widehat{\Theta})|$ gives a higher precision of the estimators of the parameters.  The $D$- optimality criterion is defined as $minimizing$ $|I^{-1}(\widehat{\Theta})|$. 
		\subsection*{$F$-optimality}
		This criterion is based on the trace of the first order approximation of the Fisher information matrix of the MLEs. The trace of $I(\widehat{\Theta})$ equals to the sum of the diagonal elements of $I(\widehat{\Theta})$. The $F$- optimality criterion is defined as $maximizing$ $trace$ $I(\widehat{\Theta})$. 
	\\\\
	%In the statistical literature $A$- and $D$-optimality criteria have been widely used. Here, these two criteria intend to minimize the trace and determinant of VCM, respectively. Further, $F$-optimality intends to minimize the trace of the observed Fisher information matrix of the MLEs.
	According to these optimality criteria, the corresponding optimal censoring plans have been considered in Table $\ref{T8}$ and Table $\ref{T11}$.
	\begin{table}[htbp!]
		\begin{center}
			\caption{Different optimality criterion.}
			\label{T5}
			\tabcolsep 7pt
			\small
			\scalebox{1}{
				\begin{tabular}{*{3}c*{2}{r@{}l}}
					\toprule
					\multicolumn{1}{c}{Criterion}&&& & &\multicolumn{1}{c}{Goal}  \\
					\midrule
					A-optimality &&& & & minimum trace $(I^{-1}(\widehat{\Theta}))$&\\
					D-optimality &&& & &minimum det $(I^{-1}(\widehat{\Theta}))$~~~& \\
					F-optimality &&&&& maximum  trace $(I(\widehat{\Theta}))$~~~&\\
					\bottomrule
			\end{tabular}}
		\end{center}
		Here $I^{-1}(\hat{\Theta})$ is defined in the section 2.2, and $I(\hat{\Theta})$ yields the corresponding observed Fisher information matrix.\\
		\vspace{-0.5cm}
	\end{table}

	\section{Real Data Analysis}
	In this section, two real life data sets have been analyzed to illustrate the applicability of the proposed methods.\\\\
		\textbf{Data Set I:}  
	A real life step-stress data set from Greven et. al  \cite{greven2004parametric} has been analyzed to illustrate the estimation methods developed in this paper. This data set represents $15$ fish swam initially upto $90$ minutes at a flow rate $15$ cm/sec. The time at which any fish felt fatigue and changed its position is considered as the failure time. This data set contains four stress levels and those stress levels has been considered by increasing flow rate ($5$ cm/sec) in every $20$ minute. As similar as Nassar et. al \cite{nassar2021bayesian}, we consider this data set as a simple step-stress data set by considering the first level as initial stress and merged other stress levels into one. For computational purpose, each data values have been subtracted by $50$ and divide by $150$, respectively. The transformed data set has been tabulated in Table $\ref{T6}$. \\
	
	\begin{table}[htbp!]
		\begin{center}
			\caption{Transformed data set. }
			\label{T6}
			\tabcolsep 7pt
			\small
			\scalebox{0.95}{
				\begin{tabular}{*{3}c*{2}{r@{}l}}
					\toprule
					\multicolumn{1}{c}{Stress level} & \multicolumn{1}{c}{Failure times}  \\
					\midrule
					$s_{1}$ & 0.2733, 0.2867, 0.2933, 0.3213 \\
					$s_{2}$ & 0.4387, 0.4400, 0.4433, 0.4483, 0.5117, 0.5167, 0.6955, 0.7300, 0.7600, 0.8933, 0.9222. \\
					\midrule
			\end{tabular}}
		\end{center}
		\vspace{-0.5cm}
	\end{table}

	%	In this section, we consider a real data set from Lee and Wang \cite{lee2003statistical} to illustrate the estimation methods developed in this paper. The data set contains   remission time (in months) of a random sample of $124$ bladder cancer patients and the corresponding failure times are given in Table $\ref{T7}$.
	% which is given below :\\[0.2cm]
	%-----------------------------------------------------------------------------------------------------------------------\\
	%0.08, 2.09,
	%2.73, 3.48, 4.87, 6.94, 8.66, 13.11, 23.63, 0.20, 2.22,
	%3.52, 4.98, 6.99, 9.02, 13.29, 0.40, 2.26, 3.57, 5.06,
	%7.09, 9.22, 13.80, 25.74, 0.50, 2.46, 3.64, 5.09, 7.26,
	%9.47, 14.24, 25.82, 0.51, 2.54, 3.70, 5.17, 7.28, 9.74,
	%14.76, 26.31, 0.81, 2.62, 3.82, 5.32, 7.32, 10.06,
	%14.77, 32.15, 2.64, 3.88, 5.32, 7.39, 10.34, 14.83,
	%34.26, 0.90, 2.69, 4.18, 5.34, 7.59, 10.66, 15.96,
	%36.66, 1.05, 2.69, 4.23,
	%5.41, 7.62, 10.75, 15.62, 43.01, 1.19, 2.75, 4.26, 5.41,
	%7.63, 17.12, 46.12, 1.26, 2.83, 4.33, 5.49, 7.66, 11.25,
	%17.14, 79.05, 1.35, 2.87, 5.62, 7.87, 11.64, 17.36,
	%1.40, 3.02, 4.34, 5.71, 7.93, 11.79, 18.10, 1.46, 4.40, 5.85, 8.26, 11.98, 19.13, 1.76, 3.25, 4.50,
	%6.25, 8.37, 12.02, 2.02, 3.31, 4.51, 6.54, 8.53, 12.03,
	%20.28, 2.02, 3.36, 6.93, 8.65, 12.63, 22.69.\\
	%-----------------------------------------------------------------------------------------------------------------------\\[0.2 cm]
	To check the goodness-of-fit of this data to the Gumbel Type-II distribution, K-S test has been employed. From this test, we observe the K-S distance is $0.2667$ and the corresponding $p$-value is $0.6781$. The MLEs of the model parameters for complete real data set are $\widehat{\alpha}=2.8443$ and $\widehat{\lambda}=0.0762$. For this data set, the K-S distance and $p$-value (in bracket) corresponding to Weibull and exponential distributions are $0.9833 (2.2 \times 10^{-6})$ and $0.4019 (0.0106)$, respectively. This represents that the given data set fits the Gumbel Type-II distribution better than Weibull and exponential distributions.  Also, for the purpose of goodness-of-fit test, different plots are considered in Figure $1$ and Figure $2$. Figure $1$ represents the comparison between the theoretical CDF of the Gumbel Type-II distribution and the empirical CDF, P-P plot and Q-Q plot of the given real data set. If $(X_{1},\cdots,X_{n})$ are $n$ number of i.i.d. random variables with CDF $F(t)$, then the empirical CDF (ECDF) is given as $F_{n}(t)= \frac{1}{n}\sum_{i=1}^{n} I_{X_{i} \leq t}$, where $I_{W}$ denotes the indicator of the event $W$. Here Q-Q plot represents the points $(F^{-1}(i/(n+1)),x_{(i)})$, where $x_{(i)}$ denotes the ordered data for $i=1,\cdots,n$. Then Figure $2$ represents the comparison between the theoretical density of the Gumbel Type-II distribution and the histogram and box-plot of given real data. From the box-plot, it can be concluded that the given distribution is right-skewed.

	Different simple step-stress samples based Type-II censoring scheme are considered by using different values of $\beta$ and $r$ when $\alpha=2.8443$, $\lambda= 0.0762$ and $\tau=0.4$. In Table $\ref{T7}$ computed values of the MLEs, Bayes estimates based on SEL and LL functions, average length of ACIs, BCIs and HPD credible intervals based on the real data set are tabulated. It is observed that the Bayes estimates perform better than MLEs. Further it has been noticed that the HPD credible interval performs better than asymptotic and bootstrap confidence intervals. From Table $\ref{T8}$, we can conclude that censoring plan under consideration $\beta=0.15$ and $r=10$ is the optimal plan according to the above discussed three optimality criteria among the other considered censoring plans.
	
	\begin{figure}[htbp!]
		\begin{center}
			\subfigure[]{\label{c1}\includegraphics[width=2 in]{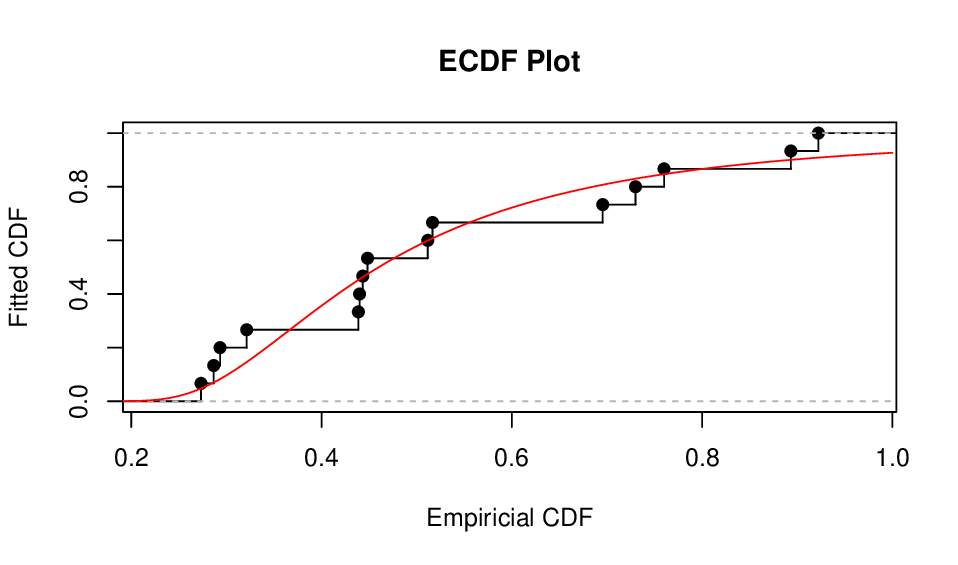}}
			\subfigure[]{\label{c1}\includegraphics[width=2 in]{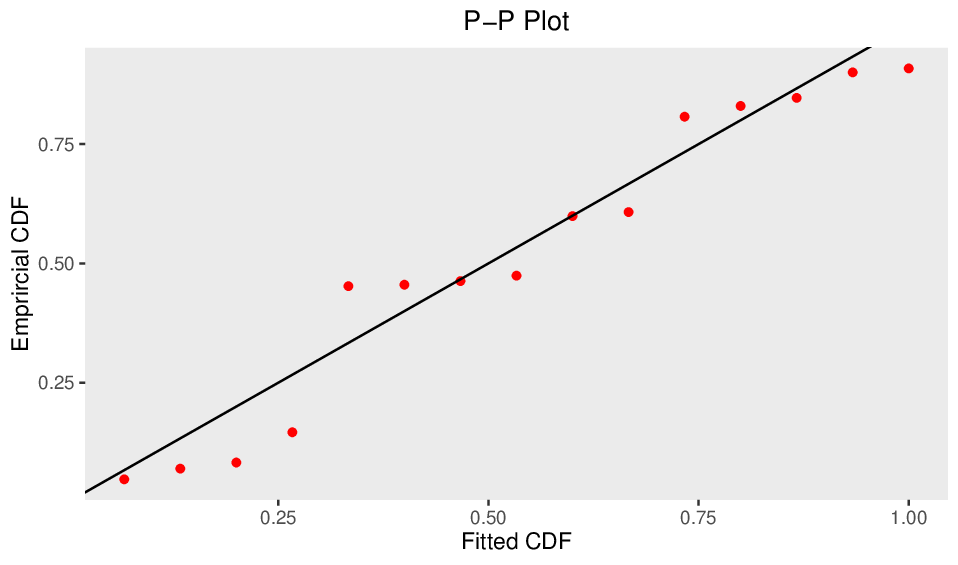}}
			\subfigure[]{\label{c1}\includegraphics[width=2 in]{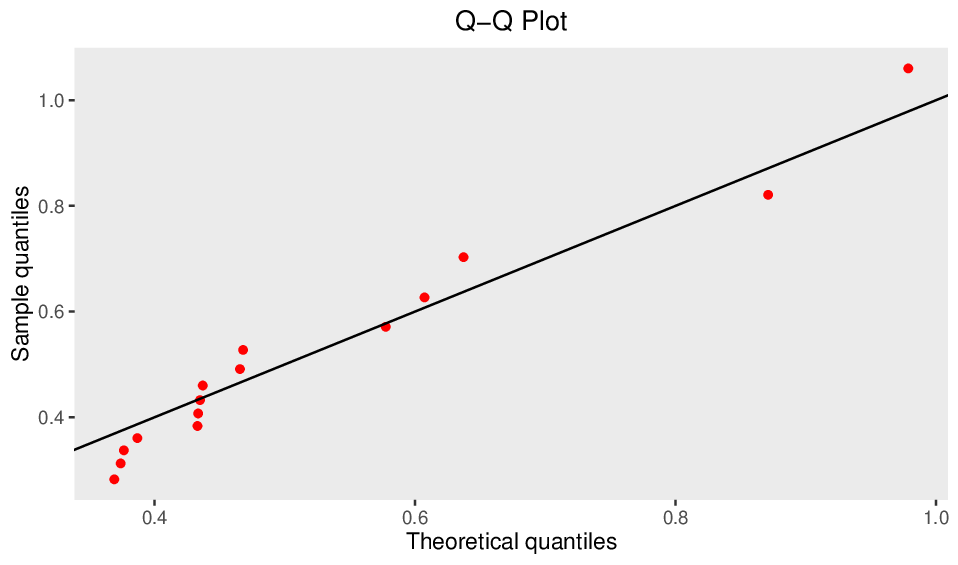}}
			\caption{(a) ECDF and CDF comparison, (b) P-P plot and (c) Q-Q plot for the Gumbel Type-II distribution fitted to given data set I.  }
		\end{center}
	\end{figure}
	
	\begin{figure}[htbp!]
		\begin{center}
			\subfigure[]{\label{c1}\includegraphics[width=3.2 in]{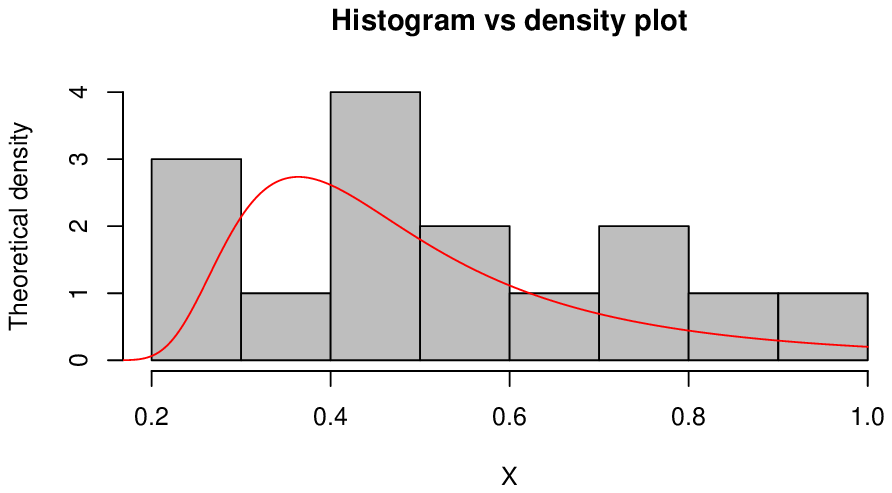}}
			\subfigure[]{\label{c1}\includegraphics[width=3.2 in]{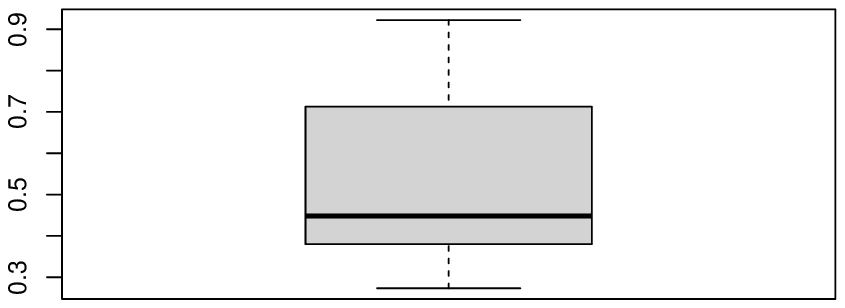}}
			\caption{(a) Histogram and theoretical density comparison and (b) Box-plot for the Gumbel Type-II distribution fitted to given data set I.  }
		\end{center}
	\end{figure}

	\begin{table}[htbp!]
		\begin{center}
			\caption{Simulation  results of classical and Bayesian estimates and length of ACIs, BCIs and HPD confidence intervals of the parameters ($\Theta$) for the Gumbel Type-II  baseline  lifetime  distribution under simple SSLT based on given real data. }
			\label{T7}
			\tabcolsep 7pt
			\small
			\scalebox{0.85}{
				\begin{tabular}{*{12}c*{11}{r@{}l}}
					\toprule
					\multicolumn{1}{c}{$\beta$} & \multicolumn{1}{c}{$r$}
					& \multicolumn{1}{c}{$\Theta$} & \multicolumn{1}{c}{MLE} & \multicolumn{1}{c}{SEL} & \multicolumn{1}{c}{LL($u=-0.05$)} & \multicolumn{1}{c}{LL($u=1$)} & \multicolumn{1}{c}{ACI} & \multicolumn{1}{c}{BCI} & \multicolumn{1}{c}{HPD} \\
					\midrule
					0.15&  8& $\alpha$& 2.7183& 2.7967& 2.7968& 2.7964& 3.7251& 5.0523& 0.4717 \\[1 mm]
					&  & $\lambda$& 0.0882& 0.0818& 0.0818& 0.0817& 0.3876& 0.2804& 0.0165 \\[1 mm]
					&  & $\beta$& 0.1008& 0.1363& 0.1363& 0.1363& 0.2917& 0.3620& 0.0343 \\ [0.2 cm] 
					& 10& $\alpha$& 2.7357& 2.8841& 2.8844& 2.8774& 3.7237& 4.5014& 0.4161   \\
					& & $\lambda$& 0.0863& 0.0796& 0.0796& 0.0796& 0.3790& 0.2246& 0.0154  \\
					& & $\beta$& 0.1196& 0.1472& 0.1472& 0.1471& 0.3096& 0.3481& 0.0307  \\
					\midrule
					0.25& 8& $\alpha$& 2.7189& 2.7933& 2.7934& 2.7922& 3.7285& 4.8818& 0.3764 \\
					& & $\lambda$&  0.0881& 0.0849& 0.0849& 0.0846& 0.3877& 0.1890& 0.0192   \\
					& & $\beta$& 0.1680& 0.1735& 0.1736& 0.1732& 0.4867& 0.3356& 0.0239  \\ [0.2 cm]
					& 10& $\alpha$& 2.7357& 2.8401&  2.8402& 2.8379& 3.7242& 3.9742& 0.3438 \\
					& & $\lambda$&  0.0863& 0.0772& 0.0772& 0.0772& 0.3791& 0.2292& 0.0185   \\
					& & $\beta$&  0.1993&  0.2116&  0.2116& 0.2115& 0.5160& 0.3114& 0.0213  \\
					\midrule
			\end{tabular}}
		\end{center}
		\vspace{-0.5cm}
	\end{table}
	
	\begin{table}[htbp!]
		\begin{center}
			\caption{Different optimality criteria for the Gumbel Type-II  baseline  lifetime  distribution under simple SSLT based on given real data. }
			\label{T8}
			\tabcolsep 7pt
			\small
			\scalebox{0.85}{
				\begin{tabular}{*{12}c*{11}{r@{}l}}
					\toprule
					\multicolumn{1}{c}{$\beta$} & \multicolumn{1}{c}{$r$}
					& \multicolumn{1}{c}{$A$-optimality}& \multicolumn{1}{c}{$D$-optimality}& \multicolumn{1}{c}{$F$-optimality}\\
					\midrule
					0.15& 8& 0.9183& 1.5899 $\times$ $10^{-6}$& 2236.2850 \\
					& 10& \textbf{0.9179} & \textbf{1.3645} $\times$ $10^{-6}$& \textbf{2377.7000} \\ 
					\midrule
					0.25& 8& 0.9299& 4.4225 $\times$ $10^{-6}$& 2032.8210 \\
					& 10& 0.9293& 3.7893 $\times$ $10^{-6}$& 2150.1210 \\
					\bottomrule
			\end{tabular}}
		\end{center}
		\vspace{-0.5cm}
	\end{table}
	\noindent \textbf{Data Set II:} A real life data set containing  the relief times of patients who received an analgesic from Gross and Clark \cite{gross1975survival} has been considered. For computational purpose the data set has been tabulated in Table $\ref{T9}$. 
		\begin{table}[htbp!]
			\begin{center}
				\caption{Transformed data set. }
				\label{T9}
				\tabcolsep 7pt
				\small
				\scalebox{0.95}{
					\begin{tabular}{*{3}c*{2}{r@{}l}}
						\toprule
						\multicolumn{1}{c}{Stress level} & \multicolumn{1}{c}{Failure times}  \\
						\midrule
						$s_{1}$ & 1.1, 1.2, 1.3, 1.4, 1.4, 1.5, 1.6, 1.6 \\
						$s_{2}$ & 1.7, 1.7, 1.7, 1.8, 1.8, 1.9, 2.0, 2.2, 2.3, 2.7, 3.0, 4.1 . \\
						\midrule
				\end{tabular}}
			\end{center}
			\vspace{-0.5cm}
		\end{table}
		To check the goodness-of-fit of this data to the Gumbel Type-II distribution, K-S test has been employed. From this test, we observe the K-S distance is $0.1016$ and the corresponding $p$-value is $0.9855$. The MLEs of the model parameters for complete real data set are $\widehat{\alpha}=4.0172$ and $\widehat{\lambda}=6.0221$. For this data set, the K-S distance and $p$-value (in bracket) corresponding to Weibull and exponential distributions are $0.1849 (0.5009)$ and $0.4395 (0.0008)$, respectively. This represents that the given data set fits the Gumbel Type-II distribution better than Weibull and exponential distributions. Also, for the purpose of goodness-of-fit test, empirical CDF plot, P-P plot, Q-Q plot, theoretical density with histogram plot and boxplot are considered in Figure $3$ and Figure $4$.
	\begin{figure}[htbp!]
		\begin{center}
			\subfigure[]{\label{c1}\includegraphics[width=2 in]{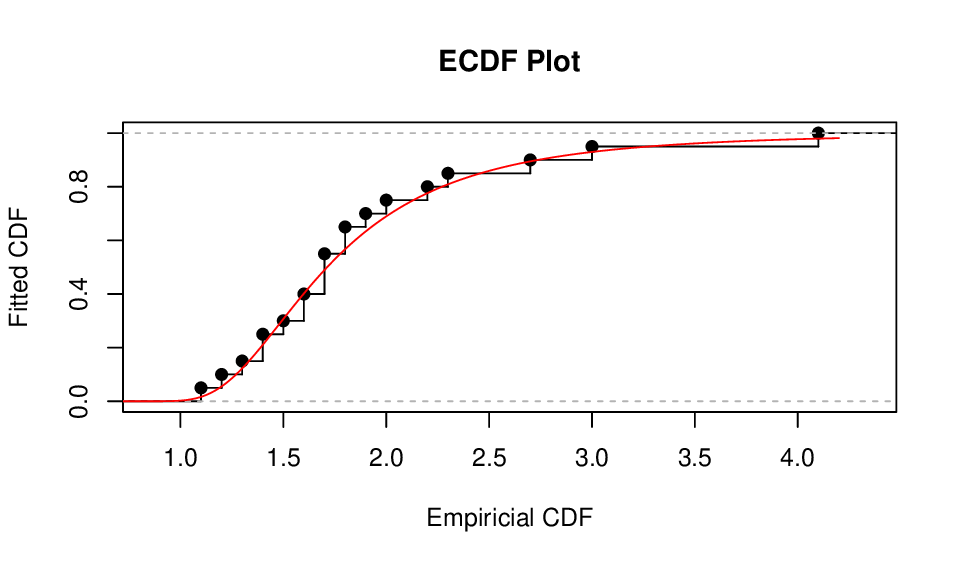}}
			\subfigure[]{\label{c1}\includegraphics[width=2 in]{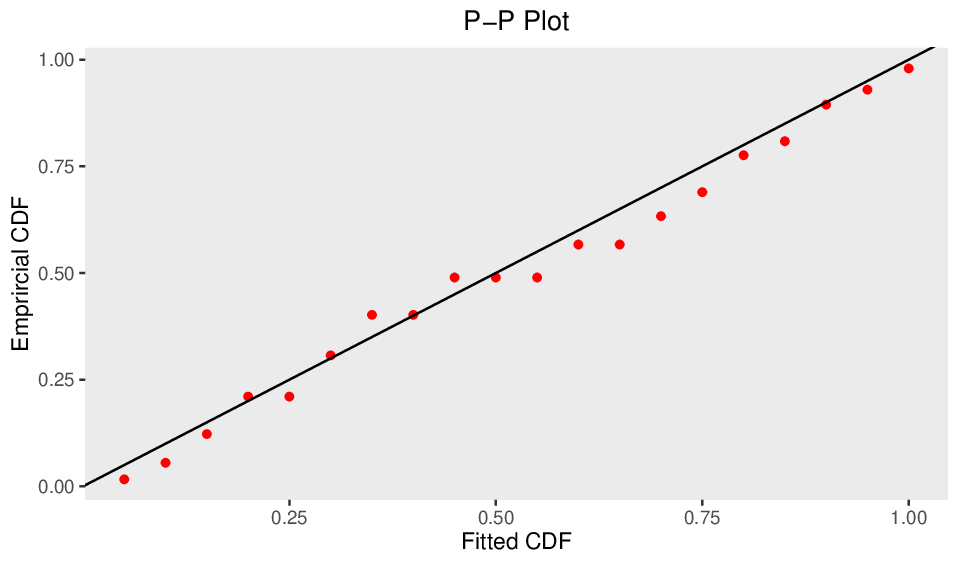}}
			\subfigure[]{\label{c1}\includegraphics[width=2 in]{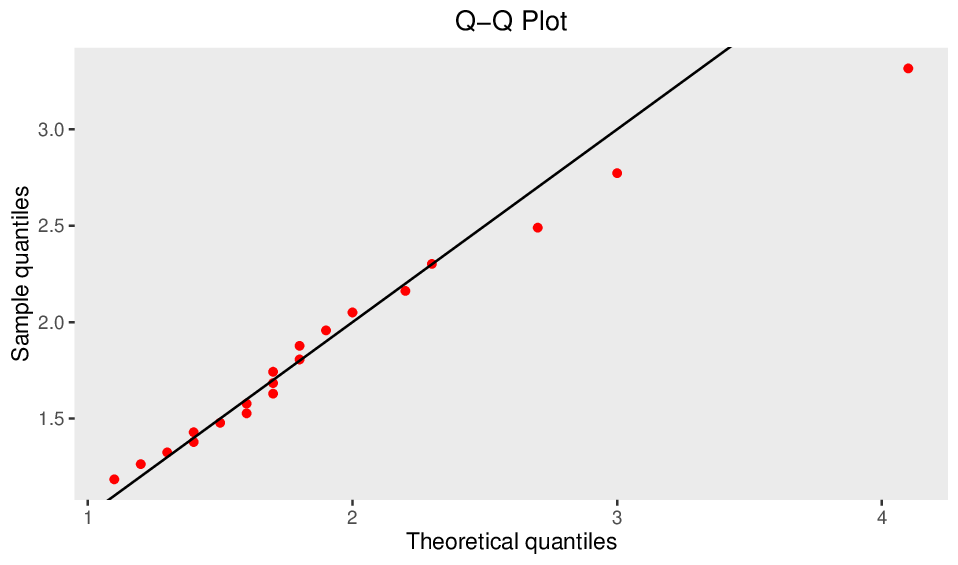}}
			\caption{(a) ECDF and CDF comparison, (b) P-P plot and (c) Q-Q plot for the Gumbel Type-II distribution fitted to given data set II.}  
		\end{center}
	\end{figure}
	\begin{figure}[htbp!]
		\begin{center}
			\subfigure[]{\label{c1}\includegraphics[width=3.2 in]{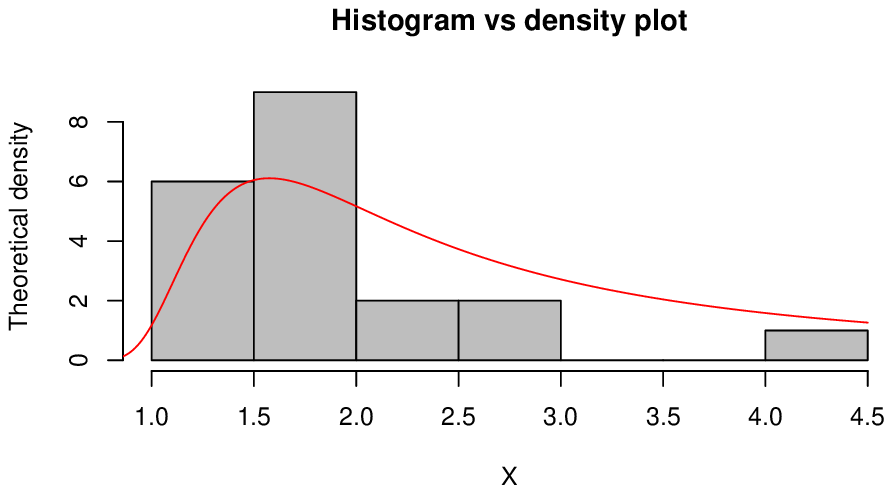}}
			\subfigure[]{\label{c1}\includegraphics[width=3.2 in]{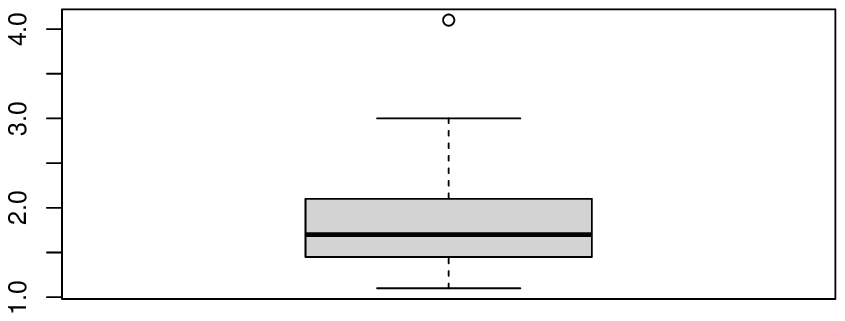}}
			\caption{(a) Histogram and theoretical density comparison and (b) Box-plot for the Gumbel Type-II distribution fitted to given data set II. }
		\end{center}
	\end{figure}
	\begin{table}[htbp!]
			\begin{center}
				\caption{Simulation  results of classical and Bayesian estimates and length of ACIs, BCIs and HPD confidence intervals of the parameters ($\Theta$) for the Gumbel Type-II  baseline  lifetime  distribution under simple SSLT based on given real data. }
				\label{T10}
				\tabcolsep 7pt
				\small
				\scalebox{0.85}{
					\begin{tabular}{*{12}c*{11}{r@{}l}}
						\toprule
						\multicolumn{1}{c}{$\beta$} & \multicolumn{1}{c}{$r$}
						& \multicolumn{1}{c}{$\Theta$} & \multicolumn{1}{c}{MLE} & \multicolumn{1}{c}{SEL} & \multicolumn{1}{c}{LL($u=-0.05$)} & \multicolumn{1}{c}{LL($u=1$)} & \multicolumn{1}{c}{ACI} & \multicolumn{1}{c}{BCI} & \multicolumn{1}{c}{HPD} \\
						\midrule
						0.25&  15& $\alpha$& 4.2428& 4.1892& 4.1655& 4.1428& 1.6211& 2.0258& 0.9565 \\[1 mm]
						&  & $\lambda$& 6.4260& 6.1472& 6.1025& 6.0883& 3.4722& 4.5213& 1.9156 \\[1 mm]
						&  & $\beta$& 0.2149& 0.2327& 0.2364& 0.2401& 0.1453& 0.1581& 0.0834 \\ [0.2 cm] 
						& 17& $\alpha$& 4.2309& 4.1563& 4.1253& 4.1198& 1.5639& 1.9857& 0.9046   \\
						& & $\lambda$& 6.4106& 6.1293& 6.0934& 6.0852& 3.4487& 4.4651& 1.8427  \\
						& & $\beta$& 0.2208& 0.2376& 0.2401& 0.2423& 0.1380& 0.1435& 0.0798  \\
						\midrule
						0.35& 15& $\alpha$& 4.2426& 4.1795& 4.1592& 4.1405& 1.6185& 1.9861& 0.9486 \\
						& & $\lambda$&  6.4266& 6.1356& 6.1008& 6.0812& 3.4650& 4.4894& 1.9052   \\
						& & $\beta$& 0.3052& 0.3214& 0.3259& 0.3295& 0.1859& 0.1921& 0.0936  \\ [0.2 cm]
						& 17& $\alpha$&  4.2148& 4.1239& 4.1052& 4.1016& 1.5699& 1.9523& 0.9172 \\
						& & $\lambda$&  6.3736& 6.1208& 6.0895& 6.0764& 3.3568& 4.3672& 1.8785   \\
						& & $\beta$&  0.3104&  0.3265&  0.3287& 0.3314& 0.1786& 0.1897& 0.0874  \\
						\midrule
				\end{tabular}}
		\end{center}
		\vspace{-0.5cm}
	\end{table}
	
	\begin{table}[htbp!]
			\begin{center}
				\caption{Different optimality criteria for the Gumbel Type-II  baseline  lifetime  distribution under simple SSLT based on given real data. }
				\label{T11}
				\tabcolsep 7pt
				\small
				\scalebox{0.85}{
					\begin{tabular}{*{12}c*{11}{r@{}l}}
						\toprule
						\multicolumn{1}{c}{$\beta$} & \multicolumn{1}{c}{$r$}
						& \multicolumn{1}{c}{$A$-optimality}& \multicolumn{1}{c}{$D$-optimality}& \multicolumn{1}{c}{$F$-optimality}\\
						\midrule
						0.25& 15& 5.7792& 0.1665 & 12.3242 \\
						& 17& \textbf{5.5784} & \textbf{0.0241} & \textbf{57.9969} \\ 
						\midrule
						0.35& 15& 5.8592& 0.2752 & 9.9783 \\
						& 17& 5.7181& 0.0646 & 24.9099 \\
						\bottomrule
				\end{tabular}}
		\end{center}
		\vspace{-0.5cm}
	\end{table}
	
	Different simple step-stress samples based Type-II censoring scheme are considered by using different values of $\beta$ and $r$ when $\alpha=4.0172$, $\lambda= 6.0221$ and $\tau=2$. In Table $\ref{T10}$ computed values of the MLEs, Bayes estimates based on SEL and LL functions, average length of ACIs, BCIs and HPD credible intervals based on the real data set are tabulated. It is observed that the Bayes estimates perform better than MLEs. Further it has been noticed that the HPD credible interval performs better than asymptotic and bootstrap confidence intervals. From Table $\ref{T11}$, we can conclude that censoring plan under consideration $\beta=0.25$ and $r=17$ is the optimal plan according to the above discussed three optimality criteria among the other considered censoring plans.

	\section{Conclusion}
	In this paper, we obtained  estimates of the unknown model parameters of the Gumbel Type-II distribution under both classical and the Bayesian approaches using TRV modeling for simple SSLT. It has been observed that the MLEs can not be obtained explicitly for all the unknown parameters. Therefore, we used the Newton-Raphson iterative method to compute MLEs by using $R$ software. Also, we obtained  the Bayes estimates based on the symmetric and asymmetric loss functions under the assumption of independent priors. A Monte Carlo simulation study is performed to compare the performance of the estimates in terms of the average values and MSEs. It has been noticed that the Bayes estimates under LL function perform better than the other point estimates. The asymptotic confidence intervals, bootstrap confidence intervals, and HPD credible intervals are also obtained. It is noticed that the HPD credible intervals perform better than other confidence intervals in terms of average width of the intervals. Further, two real life data sets are considered for illustrative purposes. An optimal censoring plan has been suggested by using different optimality criteria.\\
	\\
	\textbf{Acknowledgement:} The authors would like to thank the Editor in Chief, an Associate Editor and two anonymous reviewers for their positive remarks and useful comments.
	The author S. Dutta, thanks the Council of Scientific and Industrial Research (C.S.I.R.
	Grant No. 09/983(0038)/2019-EMR-I), India, for the financial assistantship received to carry out this
	research work. The first and third authors thank the research facilities received from the Department of Mathematics, National Institute of Technology Rourkela, India.
	\\
	\\
	{\bf The authors declare that they do not have any conflict of interests.}
	
	\bibliographystyle{ieeetr}
	\bibliography{myref}
\end{document}